\documentclass[sn-basic]{sn-jnl}

\jyear{2021}%

\theoremstyle{thmstyleone}%
\theoremstyle{thmstyletwo}%

\theoremstyle{thmstylethree}%

\usepackage{lineno,hyperref}
\usepackage{booktabs}
\usepackage{multirow}
\usepackage{color}
\usepackage{colortbl}
\usepackage{xspace}
\usepackage{amsmath}
\usepackage{rotating}
\usepackage{graphicx}
\usepackage{xcolor}
\usepackage[export]{adjustbox}

\newcommand{\modA}{\textsc{m-thumbs}\xspace}
\newcommand{\modB}{\textsc{m-aspects}\xspace}
\newcommand{\modC}{\textsc{m-summary}\xspace}
\newcommand{\modelChen}{\textsc{m-opinions}\xspace}
\newcommand{\modD}{\textsc{m-reviews}\xspace}

\begin{document}

\title[Justification of Recommender Systems Results: A Service-based Approach]{Justification of Recommender Systems Results: A Service-based Approach}


\author*[1]{\fnm{Noemi} \sur{Mauro}}\email{noemi.mauro@unito.it}

\author[1]{\fnm{Zhongli Filippo} \sur{Hu}}\email{zhonglifilippo.hu@unito.it}

\author[1]{\fnm{Liliana} \sur{Ardissono}}\email{liliana.ardissono@unito.it}

\affil*[1]{\orgdiv{Computer Science Department}, \orgname{University of Torino}, \orgaddress{\street{Corso Svizzera 185}, \city{Torino}, \postcode{10149},  \country{Italy}}}




\abstract{With the increasing demand for predictable and accountable Artificial Intelligence, the ability to explain or justify recommender systems results by specifying how items are suggested, or why they are relevant, has become a primary goal.
However, current models do not explicitly represent the services and actors that the user might encounter during the overall interaction with an item, from its selection to its usage. Thus, they cannot assess their impact on the user's experience.
To address this issue, we propose a novel justification approach that uses service models to (i) extract experience data from reviews concerning all the stages of interaction with items, at different granularity levels, and (ii) organize the justification of recommendations around those stages. 
In a user study, we compared our approach with baselines reflecting the state of the art in the justification of recommender systems results. 
The participants evaluated the \textit{Perceived User Awareness Support} provided by our service-based justification models higher than the one offered by the baselines. Moreover, our models received higher \textit{Interface Adequacy} and \textit{Satisfaction} evaluations by users having different levels of Curiosity or low Need for Cognition (NfC). Differently, high NfC participants preferred a direct inspection of item reviews.
These findings encourage the adoption of service models to justify recommender systems results but suggest the investigation of personalization strategies to suit diverse interaction needs.}

\maketitle


\section{Introduction}
\label{sec:introduction}
The demand for predictable and accountable Artificial Intelligence (AI) grows as tasks with higher sensitivity and social impact are more commonly entrusted to AI services \citep{Mohseni-etal:21} and important decisions are delegated to them \citep{Springer-Whittaker:19}. Moreover, with the introduction of the European General Data Protection Regulation \citep{GDPR}, which prescribes the user's ``right to obtain meaningful information about the logic involved'' (right to explanation), transparency has become a mandatory condition for all intelligent systems. 

Recommender systems \citep{Ricci-etal:22} are the mainstream AI technique that supports information filtering. In several application domains, such as information exploration and e-commerce, they prevent overloading users with the plethora of available options to choose from. 
Thus, the ability to properly explain or justify the recommendations has become a primary goal \citep{DiNoia-etal:22}.
A large amount of work focuses on shading light on the internal system behavior to increase recommender systems transparency by explaining {\em how} they suggest items \citep{Nunes-Jannach:17,Tintarev-Masthoff:12,Tintarev-Masthoff:22,Jannach-etal:19}. This aspect has been positively associated with the acceptance of results because it manifests the logic behind them \citep{Cramer-etal:08,Pu-Chen:07,Tintarev-Masthoff:22}. Moreover, some justification models have been developed to face the challenges of black-box models, whose internal behavior is difficult to interpret, by specifying {\em why} the system provides certain results \citep{Musto-etal:20,Ni-etal:19}. However, both types of approaches are unaware of the service model that determines the interaction with an item, from its selection to its usage. Thus, they cannot take the overall consumer experience into account in the presentation of recommendations. 

In service modeling research, \cite{sdt11} point out that items are complex entities whose fruition might involve stages of interaction with multiple services and actors that jointly impact customer experience. For instance, in the services related to the circular economy, such as home-booking, the offered value goes beyond the characteristics of the homes and includes getting in contact, or sharing spaces with their hosts. This implies different attitudes toward renting rooms or complete apartments \citep{Lee:22}. Moreover, people can be exposed to amateur providers who might offer a low quality of service level \citep{Yi-etal:20}. 
As exogenous risk factors can impact the overall interaction with items, recommender systems should explain, or justify, their own suggestions by providing users with a holistic view of items. 

Review-based recommender systems \citep{Chen-etal:15,Rubio-etal:19} recognize the importance of consumer feedback to extract experience data about items but they overlook the structure of the underlying service. Therefore, they provide users with item-centric information that partially supports decision-making.

We investigate a service-based information presentation approach to make users aware of the overall experience they should expect when selecting items. In this context, we pursue the justification of recommender systems results because it is agnostic with respect to the applied algorithms but can be exploited to enhance users' awareness of the pros and cons of the items suggested by the recommender.
\cite{Mauro-etal:22b} tested the recommendation performance of a few service-aware recommender systems that leverage consumer feedback to extract coarse-grained experience evaluation dimensions of items. Those dimensions guide (i) the rating estimation, (ii) a visual summarization of the sentiment emerging from the reviews, and (iii) the indexing of reviews by evaluation dimension.
However, that work does not support the presentation of item aspects, nor the justification of recommendation results, which we investigate in the present paper.
Specifically, here we advance that work in different ways. First, we use the Service Blueprints \citep{Bitner-etal:08} to define a more detailed service model that describes the stages of interaction with items. Second, we use that model to extract experience data from reviews concerning the stages of interaction with items at two granularity levels related to coarse-grained and fine-grained evaluation dimensions. Third, we use the service model to organize the item aspects in the justification of recommendations around these two types of evaluation dimensions and we associate aspects to the service stages that the user is expected to engage in.
With respect to the explanatory aims that \cite{Tintarev-Masthoff:12} identified, we are interested in evaluating the impact of service-based models on effectiveness and satisfaction. Specifically, we pose the following research questions:
\begin{itemize}
    \item 
    RQ1: {\em How does a service-based justification of recommendations impact the user's awareness about items and her/his confidence in evaluating them?}
    \item 
    RQ2:
    {\em How does a service-based justification of recommendations impact the user's satisfaction with the presentation of information about items?}
\end{itemize}
We developed two service-based justification models that use coarse-grained and fine-grained evaluation dimensions of the experience with items to present the key aspects emerging from consumer feedback. These models organize the access to item aspects differently. However, both of them support an incremental information exploration to enable the inspection of data depending on diverse interests in the evaluation dimensions. We defined these dimensions by applying the Service Blueprint model \citep{Bitner-etal:08}.

In a user study involving 59 participants, we compared our justification models with an approach similar to \citep{Musto-etal:20}, an aspect-based comparison of items like \citep{Chen-etal:14}, and a feature-based presentation of items and reviews inspired by standard e-commerce web sites. We implemented all these models in a test application that supported the user study. All the participants evaluated the \textit{Perceived User Awareness Support} provided by our models higher than the one offered by the baselines.
Moreover, the people having high or low values of the Curiosity trait \citep{Kashdan:09}, and those having low Need for Cognition (NfC) \citep{Coelho-etal:20}, evaluated the \textit{Interface Adequacy} and \textit{Satisfaction} of our models higher than the baselines. Differently, high NfC participants preferred the direct inspection of item reviews.
These findings encourage the adoption of service models in the justification of recommender systems results but suggest to investigate personalization to suit diverse interaction styles.

In the following, we introduce the Service Blueprints and the related work (Sections \ref{sec:background} and \ref{sec:related}). Then, we describe the dataset we used and the justification models (Sections \ref{sec:data} and \ref{sec:models}). In Sections \ref{sec:preliminary}, \ref{sec:study} and \ref{sec:experiments}, we present some preliminary findings, the user study and its results. Section \ref{sec:discussion} discusses the results and Section \ref{sec:limitations} reports limitations and future work. Finally, we discuss the ethical issues and we conclude the article (Sections \ref{sec:ethics} and \ref{sec:conclusions}).

\begin{sidewaysfigure}
\includegraphics[width=1\columnwidth]{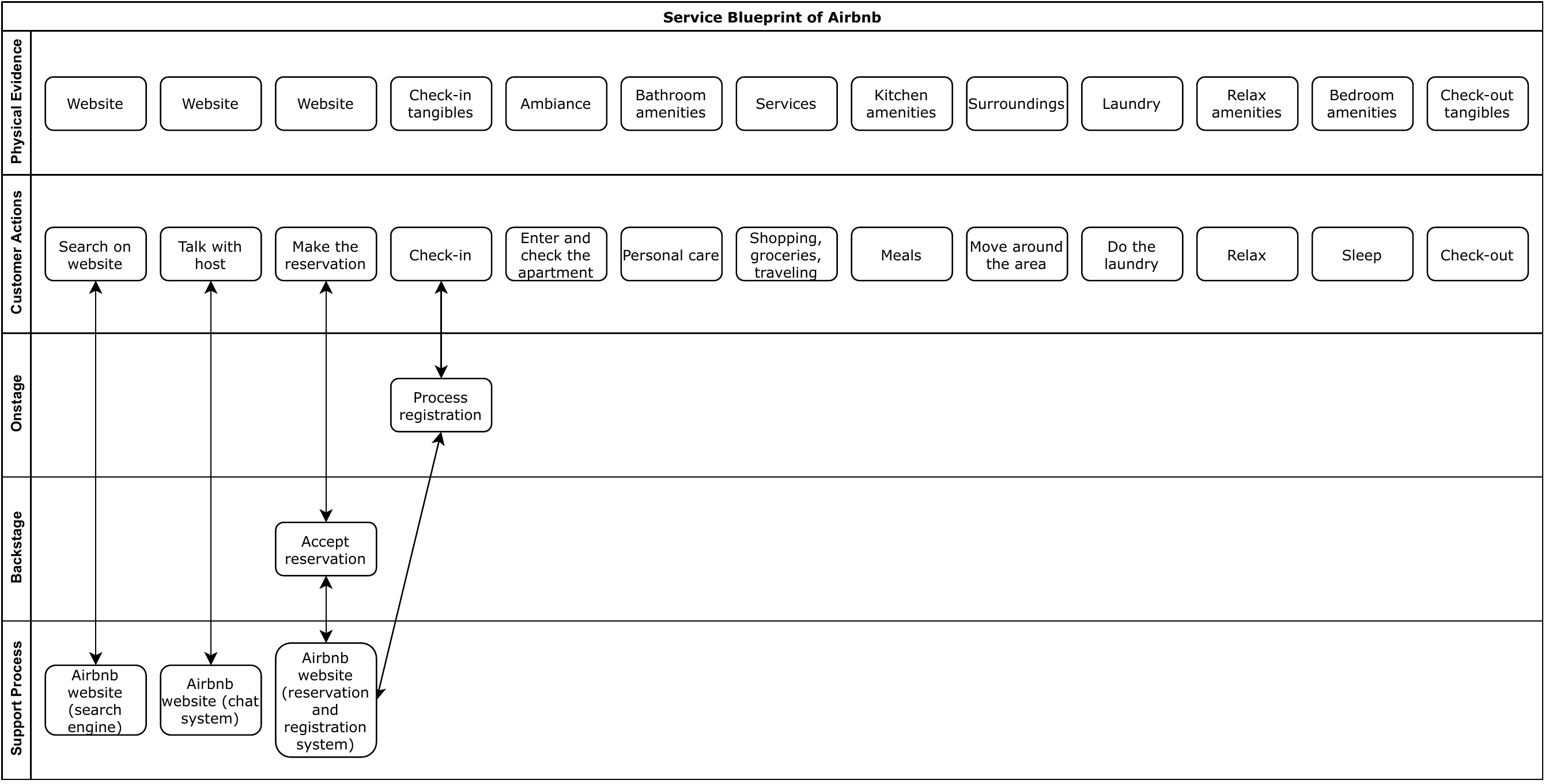}
\caption{Service Blueprint we defined to describe the home-booking domain (Airbnb).}
\label{fig:blueprint}
\end{sidewaysfigure}  

\section{Background on service blueprints}
\label{sec:background}
Service Blueprints \citep{Bitner-etal:08} are visual models that support the design and development of products and services by focusing on customers' viewpoint during the stages a person engages in, from the start point (e.g., enter website or shop) to the end one (customer care). 
They are largely used in service and product modeling and there are many examples, especially for e-commerce. \cite{Bitner-etal:08} provides a sample blueprint that describes consumer experience in the hotel booking domain; we used that example to specify the home-renting process with \cite{Airbnb}, shown in
Fig. \ref{fig:blueprint}:
\begin{itemize}
\item
The {\em Physical evidence} includes the tangibles that the customer comes in contact with. For instance, in a home-booking service, this component represents the website of the home-booking platform, the check-in tangibles (e.g., the presence of key lock-boxes, keypad or smart locks), the services and the amenities concerning the rooms, the surroundings, and so forth.
\item
The {\em Customer actions} include the actions that the guest carries out during service fruition. For instance, the reservation, the arrival at the home, the activities related with personal care and home management.
\item
The {\em Onstage/visible contact employee actions} are the actions that service providers perform while they interact with the customer, such as the processing of the registration in the home at check-in time.
\item 
Other layers represent backstage contact employee actions and support processes but we omit them because they are not relevant to the customer's direct experience.

\end{itemize}
We use the domain model built through Service Blueprints (i) to steer the analysis of item reviews by organizing feedback around the specified service stages, and (ii) to structure the presentation of item aspects according to the expected user experience during such stages.

\section{Related work}
\label{sec:related}
\subsection{Premises}
The increasingly common deployment of intelligent systems in services has shown that ``when decisions are taken or suggested by automated systems, it is essential for practical, social, and—with increasing frequency—legal reasons that an explanation can be provided to users, developers, and regulators'' \citep{Confalonieri-etal:21}. 
Early recommender systems were black-boxes focused on algorithmic performance. However, \cite{Herlocker-etal:00} recognized the importance of explaining their results to enhance users' acceptance and trust. Since then, work has been done to improve the understanding of personalized recommendations \citep{Nunes-Jannach:17,Tintarev-Masthoff:12,Tintarev-Masthoff:22,Jannach-etal:19}. The approaches developed so far can be classified in (i) those that explain the system's behavior \citep{Conati-etal:21,Kouki-etal:20}, (ii) those that fuse recommendation and explanation in the same process \citep{Dong-Smyth:17,Lu-etal:18b,Rana-etal:22}, and (iii) those that provide post-hoc justifications of the suggestions \citep{Musto-etal:20,Ni-etal:19}. While the last approach is agnostic to the recommendation algorithm, the first two are tightly coupled to it.
We aim at advancing the justification approach because it can be implemented on top of different recommendation algorithms to enhance users' awareness of the suggested items.

\subsection{Explanation and justification techniques}
Different techniques support the understanding of recommendations:
\begin{itemize}
    \item 
    The systems based on a single algorithm typically explain their results in terms of inference traces \citep{Herlocker-etal:00,Nunes-Jannach:17}. Some aspect-based recommender systems highlight the features of items which match, or mismatch, the user's preferences \citep{Muhammad-etal:16}. Other ones group items on the basis of their pros and cons for the user, for easy comparison \citep{Pu-Chen:07,Chen-etal:14,Chen-Wang:17}. Some information exploration support systems explain the suggestions by visualizing the relevance of items to the keywords of the submitted search queries \citep{Chang-etal:19,diSciascio-etal:16}. Graph-based recommenders use the connections between users and items as explanations of the suggestions \citep{Amal-etal:19,Wang-etal:18,Musto-etal:19b}. Some researchers assume that item recommendation and explanation should coincide because they rely on the same logic. Thus, they fuse the two processes  \citep{Dong-Smyth:17,Lu-etal:18b}, or use the strength of the possible explanations to steer the recommendation \citep{Rana-etal:22}.
    \item
    Hybrid recommender systems explain the suggestions under multiple perspectives of relevance to represent the relative impact of the integrated systems on item rankings. For example, MyMovieFinder \citep{Loepp-etal:15} separately shows the recommenders that support a suggested item and RelevanceTuner \citep{Tsai-Brusilovsky:19} uses stackable bars to visually integrate them. TalkExplorer \citep{Vebert-etal:16} and IntersectionExplorer \citep{Cardoso-etal:19} show multiple dimensions of relevance through bidimensional graphs and grid layouts respectively. \cite{Kouki-etal:17} present the suggestions by means of Venn diagrams and \cite{Parra-Brusilovsky:15} combine Venn diagrams with color bars to distinguish the contribution to item evaluation provided by the integrated recommenders.
\end{itemize}
All these models are unaware of the service underlying items. Thus, they present item-centric data that limits users' awareness of the experience they should expect from their own selections. For instance, the quality of the post-sales customer care depends on the retailer and introduces a further dimension of choice on top of the selection of the product that fits specific needs. However, a service-agnostic analysis of item reviews might hardly reveal this type of information. 
We use service blueprinting to characterize the stages of interaction with items and the involved entities. Then, we use the resulting service model to steer the extraction, organization and presentation of item aspects around coarse-grained and fine-grained experience evaluation dimensions that the user can consider in the analysis of items. 

Commercial platforms like \cite{Airbnb} and \cite{TripAdvisor} present a summary of consumer feedback that reports the overall evaluation expressed by previous customers, a set of values associated with the main aspects of the items (e.g., cleanliness and communication for Airbnb, or cuisine and atmosphere for TripAdvisor), and the reviews.
However, they do not show the connections between aspects values and reviews, leaving the user the burden of summarizing the opinions that emerge from them. 
 
\citep{Chen-etal:14} extract item aspects from reviews and present them in quantitative and qualitative format. However, they fail to explicitly model the stages of interaction with the service behind items. Therefore, they present a flat overview of item-centric aspects that users have to interpret in terms of experience evaluation dimensions.

Some researchers discovered that users' perceptions of the explanations generated by recommender systems depend on their cognitive style \citep{Millecamp-etal:19,Millecamp-etal:20}, personality \citep{Kouki-etal:19,Millecamp-etal:20}, and domain expertise \citep{Kouki-etal:19}. 
Moreover, \cite{Millecamp-etal:22} suggested to tailor the explanations to users' personal characteristics. 
The adaptation of the justification model to the individual user is an interesting future development and is supported by our experimental results, where we noticed that users' Curiosity trait and Need for Cognition impact the perception of the justification models. See Section \ref{sec:experiments}.

\begin{table}[t]
\centering
\caption{Descriptive statistics of the filtered dataset.}
\label{tab:statistics}
\begin{tabular}{lllll}
\toprule
                      & Min & Max  & Mean  & Standard Deviation \\ \midrule
Words per review      & 1   & 1002 & 47.00 & 46.41    \\
Reviews per listing   & 1   & 648  & 20.80 & 35.96    \\
Amenities per listing & 0   & 66   & 20.98 & 7.85     \\ \bottomrule
\end{tabular}%
\end{table}

\section{Data}
\label{sec:data}
For this study we used a public dataset of Airbnb reviews concerning the homes of London city. We downloaded this dataset in January 2021 from \url{http://insideairbnb.com/get-the-data.html}.

The dataset contains information about homes (denoted as ``listings''), their administrators (``hosts'') and features (``amenities''). Moreover, it includes the reviews uploaded by the people who rented the homes (``guests''), starting from December 21st, 2009. We noticed that, in January 2020, people rented very few homes, probably because of the COVID-19 pandemic. We thus decided to filter out the reviews uploaded after the first day of that month. Then, we selected the reviews written in English and we removed the listings that did not receive any comments since 2018 to work on recently rented homes.

The filtered dataset contains 764,958 guests, 906,967 reviews and 43,604 listings, out of which we selected the homes used in our user study. Table \ref{tab:statistics} reports the descriptive statistics of that dataset.

An analysis of the reviews shows that, while some of them are extremely concise (e.g., {\em ``Amazing location!''}), other ones are fairly articulated. Reviews typically discuss features and aspects of homes such as the offered services and amenities or their cleanliness. However, they frequently also include evaluations of the hosts and surroundings, providing a broad picture of guests' renting experience.
For example: 
\begin{quote}
{\em ``A warm and private place ideal for exploring London. Location was perfect and felt very safe. We stayed with our young children and they had space to stretch out with their toys, the lift was convenient and check-in was a breeze! Very clean and comfortable, we would stay here again!''}
\end{quote}

\section{Methodology}
\label{sec:models}
The methodology we applied to develop our service-based justification models is general but we present it by referring to the home booking domain.
We first describe how we defined the evaluation dimensions of experience with items that support the organization of aspects to justify recommendations. Then, we outline the extraction and analysis of aspects from item reviews.
Finally, we present our justification models and the baselines for comparison. 

Notice that the extraction and analysis of aspects is an offline task. It should be performed once in the dataset of reviews and possibly periodically updated to take new entries into account.

\subsection{Evaluation dimensions of the experience with items}
\label{sec:dimensions}
The first step to identify the evaluation dimensions of experience is the definition of a Service Blueprint that represents users' experience with items.
Fig. \ref{fig:blueprint} shows the one we developed by extending \cite{Bitner-etal:08}'s hotel booking one with other representations of customers experience in hotel booking \citep{REN201613} and a detailed analysis of Airbnb customer's needs and preferences \citep{Cheng-Jin:19}. 

As we are interested in building a justification model for recommendations, rather than designing the complete home-booking service, we focus on the Customer Actions and Physical Evidence layers of the blueprint. 
These layers describe the typical sequence of steps the user can carry out and the tangibles (s)he can encounter. As specified in Fig. \ref{fig:blueprint}, the first tangible is the Airbnb web site that supports the interaction with the host of the home and the reservation. When reaching the home, the guest checks in and this activity might involve meeting the host (or a referent that we consider the host) to receive the keys. During the stay, the guest might be involved in various activities, such as personal care, shopping, managing meals, moving around in the neighborhood, doing the laundry and relaxing or sleeping. The last step is the check-out that, again, might involve the host.

\begin{table}[t]
\centering
\caption{Coarse-grained and fine-grained evaluation dimensions for home-booking.}
\label{tab:dimensions}
\resizebox{0.8\textwidth}{!}{%
\begin{tabular}{lll}
\toprule
Coarse-grained dimensions  & Fine-grained  dimensions & Physical Evidence \\ 
\midrule
Host appreciation  & Host & -  \\ 
\midrule
Search on website  & Website   & Website \\ 
\midrule
 & Check-in & Check-in tangibles \\
\multirow{-2}{*}{\begin{tabular}[c]{@{}l@{}}
	Check-in/Check-out\\
\end{tabular}} & Check-out & Check-out tangibles \\ 
\midrule
  & Ambiance & Ambiance \\
  & Bathroom  & Bathroom amenities \\
  & Kitchen   & Kitchen amenities \\
  & Laundry  & Laundry \\
 & Relax  & Relax amenities \\
\multirow{-7}{*}{
In apartment experience}  & Bedroom  & Bedroom amenities \\ \midrule
 & Surroundings & Surroundings \\
\multirow{-2}{*}{Surroundings} & Services  & Services \\ 
\bottomrule
\end{tabular}%
}
\end{table}

The Physical Evidence layer does not model human actors but the activities specified in the Customer Actions layer might concern different entities, including the host. As both tangibles and human actors can impact the guest's satisfaction, we add a layer to map Customer Actions to evaluation dimensions that represent the experience with the involved entities. 
To support both the summarization and a detailed organization of information about items, we define two types of experience evaluation dimensions (see Table \ref{tab:dimensions}): 
\begin{itemize}
    \item
    The {\bf fine-grained evaluation dimensions} describe consumer experience with the tangibles and actors involved in the individual customer actions. In our domain, they are the perception of the host, website of the home-booking platform, ambiance of the home, rooms, and so forth. Fine-grained dimensions also include Check-in and Check-out to represent the experience during those activities. For example, a guest might have a bad experience because the host shows up late at check-in.
    \item 
    The {\bf coarse-grained evaluation dimensions} summarize consumer experience by abstracting from individual customer actions. A coarse-grained dimension is mapped to multiple fine-grained ones. For instance, a generic ``In apartment experience" dimension can summarize the guest's experience within a home (ambiance, rooms, etc.). Similarly, check-in and check-out can be combined into a single coarse-grained dimension.
\end{itemize}
Table \ref{tab:dimensions} shows the evaluation dimensions we defined starting from the Service Blueprint of Fig. \ref{fig:blueprint}, mapped to the elements of the Physical Evidence layer.

\subsection{Extraction and organization of item aspects}
\label{sec:aspect-extraction}
We apply an extension of the approach described in \citep{Mauro-etal:21d} to extract the aspects of items emerging from the reviews. Specifically, we use the distinction between coarse-grained and fine-grained evaluation dimensions of experience to classify aspects at different granularity levels. In the following, we shortly describe how we extract and organize the aspects of an item $i$ (an individual home) starting from its reviews $REV_i$:
\begin{enumerate}
    \item
    \label{uno}
    For each $aspect$-$adjective$ pair\footnote{The extraction of aspects and adjectives can be done by applying standard NLP techniques. In our work, we exploit lemmatization, dependency parsing and the Double Propagation algorithm \citep{Qiu:2011} that is suitable to deal with datasets that have not been previously annotated because it is unsupervised.} occurring $REV_i$, we produce an
    \linebreak
    $<aspect, asp\#rev, adjective, asp\_adj\#rev, evaluation>$ tuple where:
    \begin{itemize}
        \item 
        $asp\#rev$ denotes the number of reviews $r \in REV_i$ that mention $aspect$;
        \item 
        $asp\_adj\#rev$ denotes the number of reviews $r \in REV_i$ that mention the $aspect$-$adjective$ pair;
        \item 
        $evaluation$ is the normalization in [1, 5] of the polarity of the $aspect$-$adjective$ pair. We compute the polarity as the mean value returned by the TextBlob \citep{TextBlob} and Vader \citep{Hutto-Gilbert:14} libraries.
    \end{itemize}
    These tuples summarize previous consumers' opinions about the item in a structured way by measuring how many guests mention them, and with which degree of appreciation. This differs from counting the frequency of terms and is robust to the occurrence of long reviews that repeat the same concepts several times.
    \item 
    \label{due}
    We classify the aspects extracted from $REV_i$ by fine-grained experience evaluation dimension using entity recognition (to identify references to people and places) and a set of dictionaries that collect the terms frequently used to refer to our coarse-grained dimensions. The dictionaries derive from the thesauri of \citep{Mauro-etal:20d}, which we split into subsets, each one including the terms semantically related to an individual fine-grained dimension. For example, the \texttt{Kitchen} dictionary contains ``kitchen", ``oven", ``table" and other similar keywords.
    By classifying aspects, we can associate them to the stages of interaction with items by taking the fine-grained and coarse-grained evaluation dimensions into account. This is the basis to organize information at different granularity levels, and to quantitatively summarize the related consumer feedback.

\definecolor{1}{HTML}{3FA9FC}
\definecolor{2}{HTML}{FFCE9F}
\definecolor{3}{HTML}{DAA6F7}
\definecolor{4}{HTML}{69BC4B}
\begin{table}[t]
\centering
\caption{Sample aspects extracted from the reviews of a sample Airbnb home.}
\begin{tabular}{llllll}
\toprule
aspect     & asp\#rev & adjective       & asp\_adj\#rev & evaluation & dimension    \\
\midrule
location   & 23          & great         & 6             & 4.42       & ambiance     \\

location   & 23          & excellent     & 2             & 4.57       & ambiance     \\

location   & 23          & good          & 2             & 4.14       & ambiance     \\
location   & 23          & convenient    & 1             & 3.00       & ambiance     \\
host       & 22          & great         & 7             & 4.42       & host-prop    \\
host       & 22          & friendly      & 4             & 3.87       & host-prop    \\
host       & 22          & excellent     & 2             & 4.57       & host-prop    \\
host       & 22          & lovely        & 2             & 4.09       & host-prop    \\

place      & 9           & lovely        & 3             & 4.09       & ambiance     \\
place      & 9           & great         & 2             & 4.42       & ambiance     \\

place      & 9           & airy          & 1             & 3.00       & ambiance     \\
bed        & 4           & comfortable   & 2             & 3.91       & bedroom      \\

bed        & 4           & superb        & 1             & 4.62       & bedroom      \\
restaurant & 4           & cool          & 1             & 3.67       & surroundings \\

restaurant & 4           & lovely        & 1             & 4.09       & surroundings \\
restaurant & 4           & nice          & 1             & 4.02       & surroundings\\
\bottomrule
\end{tabular} 
\label{tab:exp}
\end{table}
   
   \item
   \label{tre}
   We compute the value of each coarse-grained evaluation dimension $d$ as the weighted mean of the evaluations received by the $aspect$-$adjective$ pairs such that $aspect$ is classified in $D$. For each pair, we use the number of reviews that mention it ($asp\_adj\#rev$) as weight to tune its impact on the evaluation of $D$ coherently with the number of people who mention it. If there is no information about a dimension, its value is set to 0 that, being out of the [1, 5] range, means ``zero knowledge".
\end{enumerate}
Table \ref{tab:exp} shows the type of information that this analysis produces and aggregates data by fine-grained evaluation dimension.
Notice that, while we carry out the analysis, we index the sentences of the reviews by $aspect$-$adjective$ pair to support their retrieval for justification purposes.

\subsection{Service-based justification models}
\label{sec:service-models}
Our models summarize the main features and properties of items in the user interface and make additional data available on demand. Thus, users are free to expand the information they care about.
Fig. \ref{fig:ui} shows a portion of the user interface of the justification models we propose (the Appendix includes a larger version of this figure and of the following ones). 
We focus on the component presenting individual items: 
\begin{figure}
    \centering
    \includegraphics[width=\textwidth]{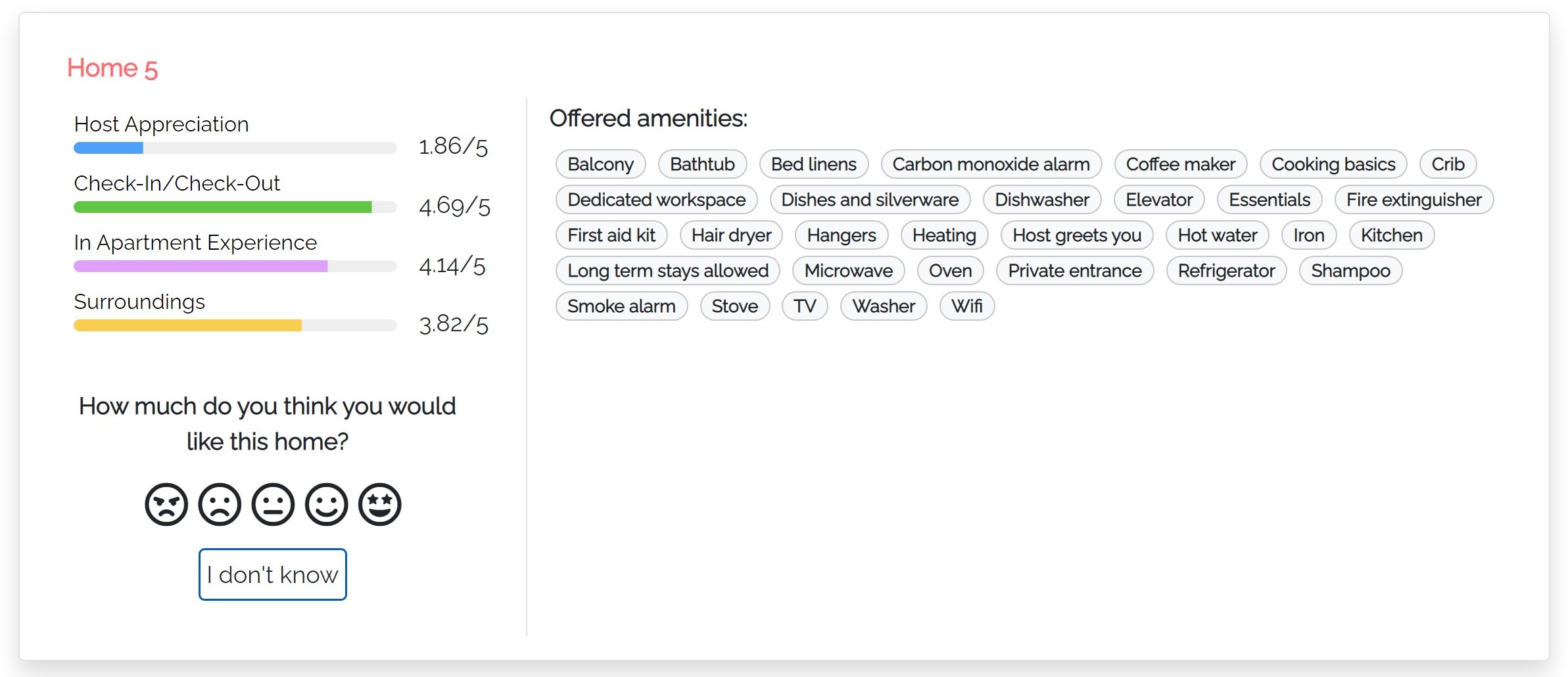}
    \caption{Portion of the user interface shared by the proposed justification models.}
    \label{fig:ui}
\end{figure}
\begin{itemize}
    \item 
    The central area shows the features of the visualized item. In our case, these are the amenities offered by the home, which typically represent binary features and thus can be shown or hidden, depending on their value. As suggested by \cite{Tintarev-Masthoff:22}, we omit data such as the price, number of rooms, pictures and name because they might influence users' evaluation of the home or their perception of the data provided by the system. For instance, some homes have long names that mention their location or the view; e.g., {\em ``Beautiful Flat - London Bridge. SE1"}, or {\em ``Clerkenwell penthouse, huge terrace''}.
    \item 
    The left area shows a set of colored bar graphs that summarize previous consumer experience with the item. Each bar corresponds to a coarse-grained experience evaluation dimension $D$, that is, \texttt{Host Appreciation}, \texttt{Check-in/Check-Out}, \texttt{In Apartment Experience} or \texttt{Surroundings} (we omit \texttt{Search on website} because we are not interested in the user's experience with the Airbnb platform). The bar shows the value of $D$, which represents the evaluation that the item has received in step \ref{tre} of Section \ref{sec:aspect-extraction}. If $D$ = 0, the name of the bar graph is displayed in light grey to denote that there is no feedback about it and distinguish the ``zero knowledge" situation from an extremely bad evaluation. The user can click on the bar graphs to receive more details about previous consumer experience with the item; see models \modA and \modB. 
    \item
    The left area also includes a rating component through which the user can evaluate the item in the [1, 5] scale, represented as a list of smilies. In this work, we do not evaluate recommendation performance but we included this component for two reasons: Firstly, we wanted to attract the user's attention to the presented data. Second, we aimed to collect some implicit feedback about her or his confidence in the evaluation of the items.
    The rating component includes an ``I don't know'' button to skip the evaluation.
\end{itemize} 
We now describe the peculiarities of our justification models, which provide different information when the user clicks the bar graphs.
    
\begin{figure}
    \centering
    \includegraphics[width=\textwidth]{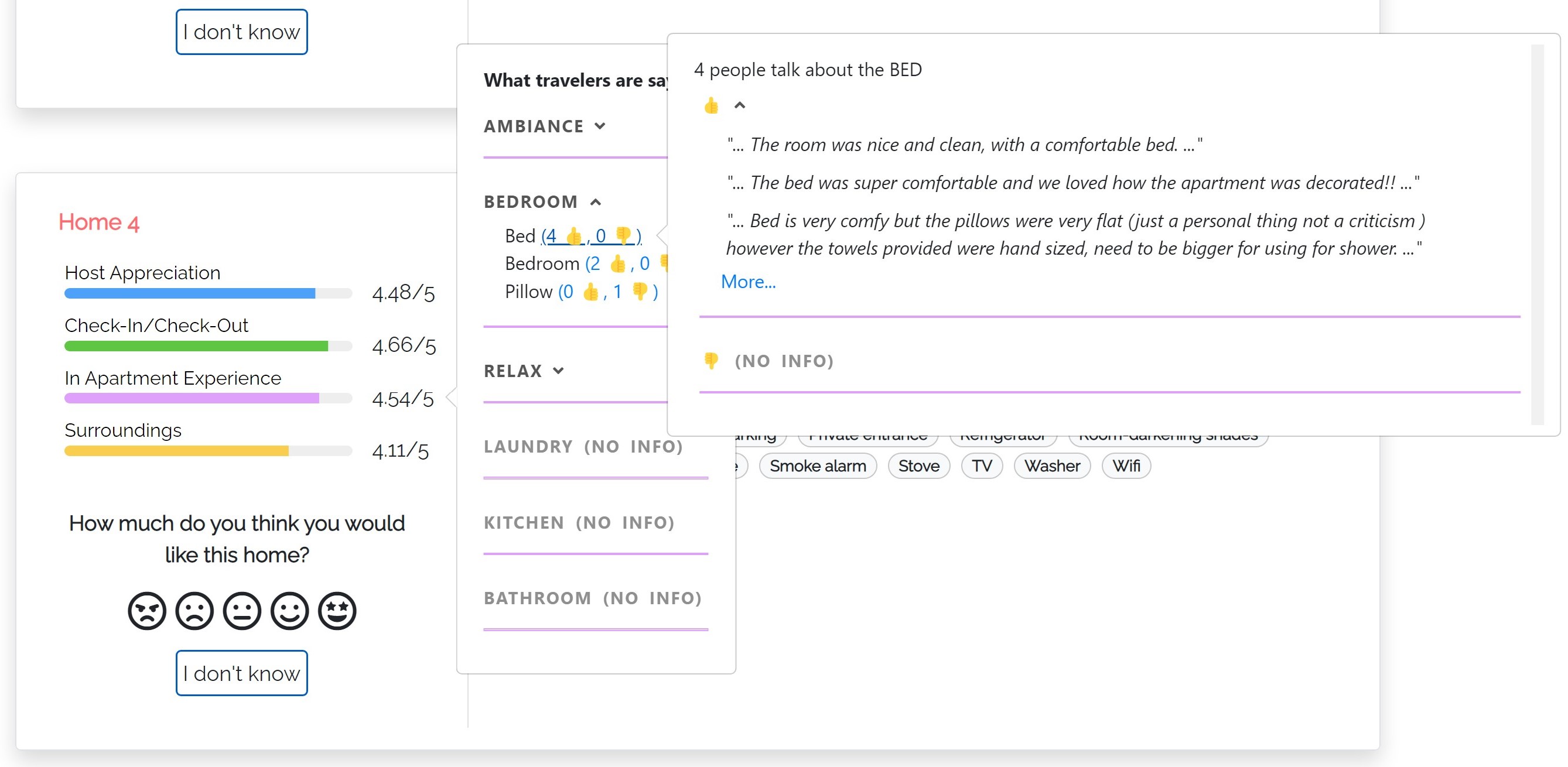}
    \caption{User interface of justification model \modA.}
    \label{fig:modelA}
\end{figure}

\subsubsection{M-THUMBS}
In this model (Fig. \ref{fig:modelA}), when the user clicks the bar of a coarse-grained dimension $D$, the user interface opens the ``What travelers are saying" component that shows the fine-grained dimensions $d \in D$. For example, ``AMBIANCE" of ``In Apartment Experience". Within the component, dimensions are sorted by the number of aspects mentioned in the reviews of the item. A dimension $d$ can be expanded to view the most relevant aspects classified in it, sorted by decreasing relevance. For clarity, we show the dimensions that have no aspects in light grey, with a (``NO INFO") tag, and they cannot be expanded. 

Here and in the following models, we compute the relevance of aspects as the number of reviews of the item that mention them; see the $asp\#rev$ field in Table \ref{tab:exp}. When the user expands a fine-grained dimension, the user interface shows a maximum of three aspects and a button to view the complete list. 
For each aspect, a thumb up/down reports the number of reviews expressing a positive/negative opinion about it, derived from the data of Table \ref{tab:exp}. 
Thumbs and numbers can be clicked to view the quotes from the reviews mentioning the aspect. 

\begin{figure}
    \centering
    \includegraphics[width=\textwidth]{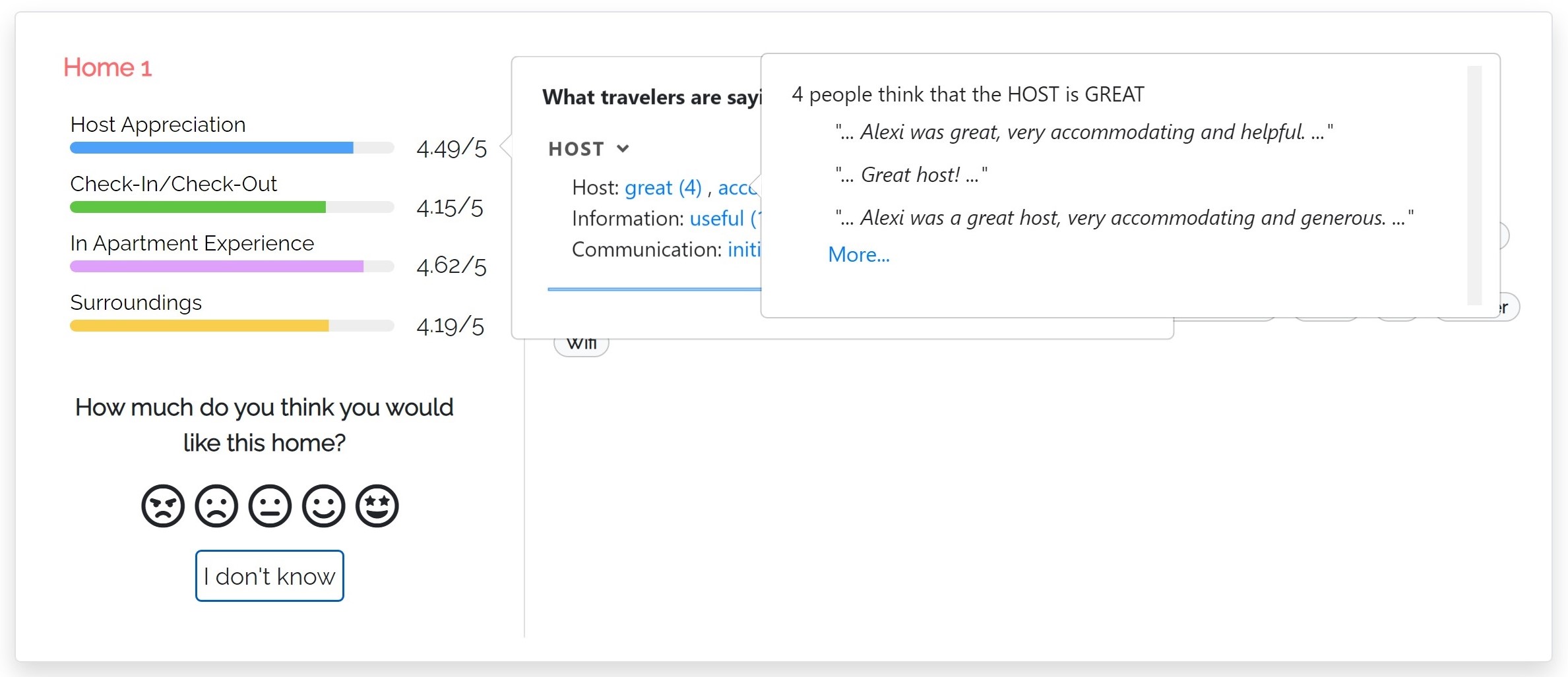}
    \caption{User interface of justification model \modB.}
    \label{fig:modelB}
\end{figure}

\subsubsection{M-ASPECTS}
In this model (Fig. \ref{fig:modelB}), the ``What travelers are saying" component is organized as in \modA. However, for each fine-grained dimension, aspects report the list of the most relevant adjectives referring to them in the reviews of the item. The relevance of an adjective corresponds to the $asp\_adj\#rev$ field of its $aspect$-$adjective$ pair in Table \ref{tab:exp}. For each adjective, the user interface reports that value and, by clicking on the term, or on the associated number, the user can view the related quotes from the reviews.

\begin{figure}
    \centering
    \includegraphics[width=\textwidth]{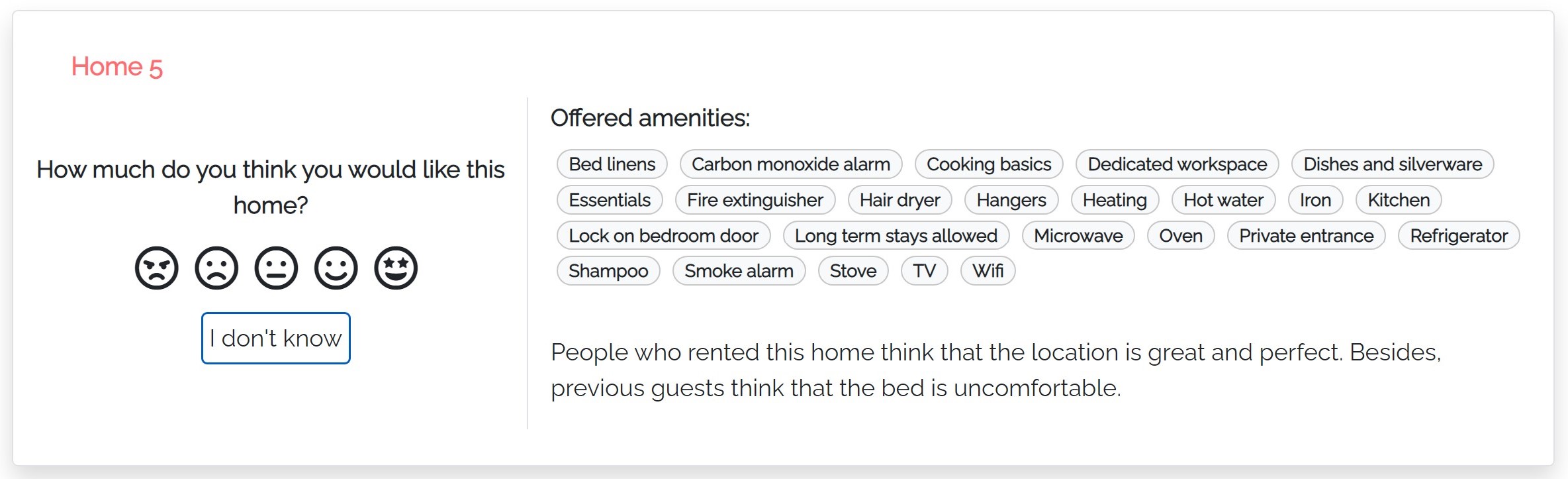}
    \caption{User interface of justification model \modC.}
    \label{fig:modelC}
\end{figure}

\begin{figure}
    \centering
    \includegraphics[width=\textwidth]{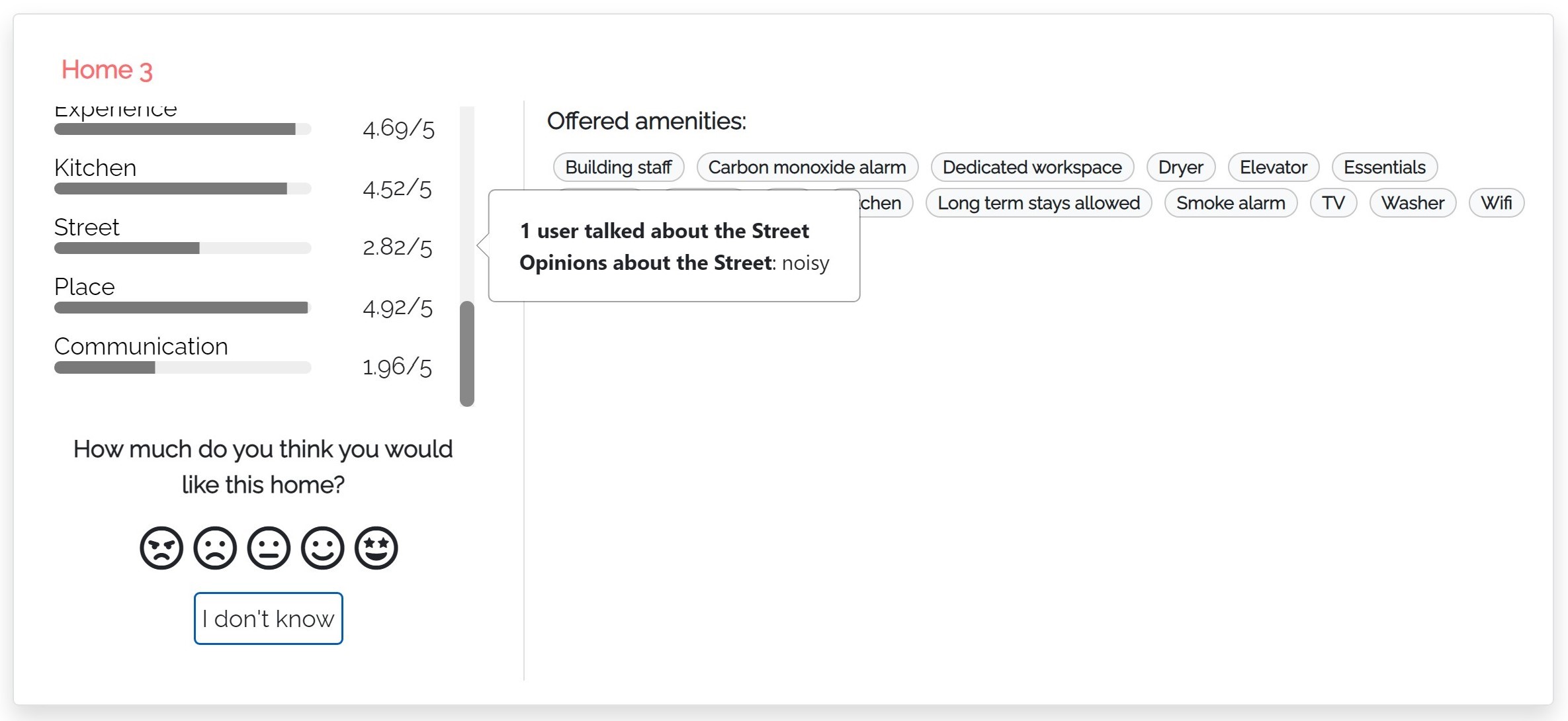}
    \caption{User interface of justification model \modelChen.}
    \label{fig:modelChen}
\end{figure}

\subsection{Baseline justification models}
\label{sec:baselines}
\subsubsection{M-SUMMARY}
In this model (Fig. \ref{fig:modelC}), below the item features, the user interface shows a summary that describes the most relevant aspects ($asp\#rev$) and adjectives ($asp\_adj\#rev$) of the item extracted from its reviews. 

Similar to \citep{Musto-etal:20}, we dynamically compose the textual paragraphs by exploiting a Backus-Naur Form (BNF) grammar that generates different types of sentences to support variability in the summaries. Moreover, we select the aspects to be included by decreasing relevance, and similar for the adjectives to be mentioned. 
However, we compute the relevance of aspects and adjectives in terms of how many reviews mention them ($asp\#rev$ and $asp\_adj\#rev$) rather than through the Kullback–Leibler divergence (KL). The reason is that KL uses a dictionary that does not suit our needs because it misses (and thus is unable to evaluate) several words that guests frequently use to express their opinions about homes.

\begin{figure}
    \centering
    \includegraphics[width=\textwidth]{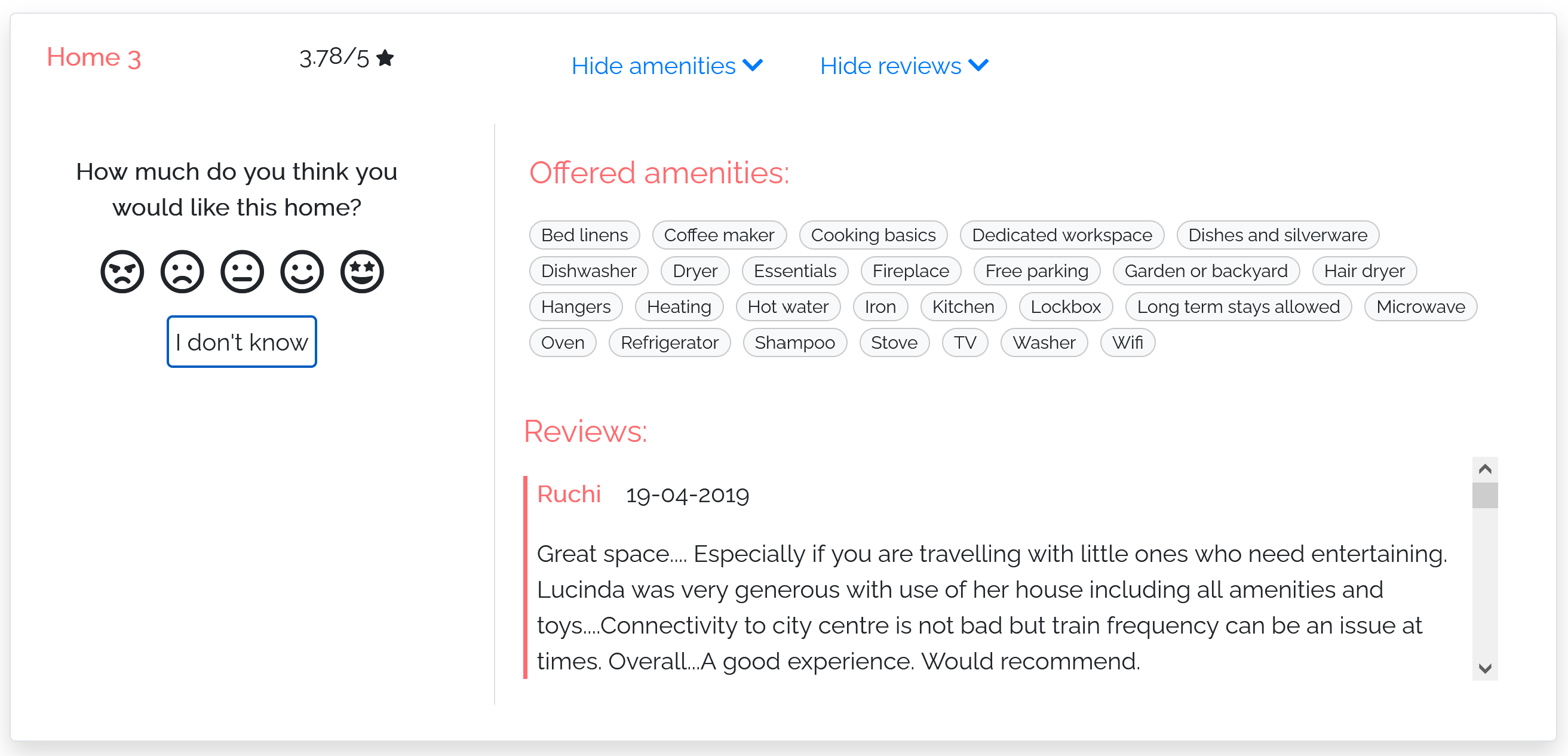}
    \caption{User interface of justification model \modD.}
    \label{fig:modelD}
\end{figure}

\subsubsection{M-OPINIONS}
In this model (Fig. \ref{fig:modelChen}), alongside the item features, the user interface shows the evaluations of the most relevant aspects of the item extracted from its reviews  ($asp\#rev$), sorted by decreasing relevance. 
Each aspect evaluation is visualized as a grey bar coupled with a numeric value in [1, 5]. By clicking the bar, the user can view a component that shows the number of guests who mentioned the aspect and the adjectives they used to qualify it. 

The value of a bar is the weighted mean of the evaluations of the related $aspect$-$adjective$ pairs ($evaluation$ in Table \ref{tab:exp}); the weight is their $asp\_adj\#rev$.

Overall, the user interface provides similar information about item aspects as \cite{Chen-etal:14}'s opinion bar chart. However, we use simple bar charts to make the visualization comparable with our service-based justification models. 

\subsubsection{M-REVIEWS}
In this model (Fig. \ref{fig:modelD}), the user interface shows the typical data provided by e-commerce platforms, and in particular by home-booking services; i.e., item features, mean rating received from consumers, and reviews. The model enables the user to hide or show the features and the reviews by clicking on the tabs available at the top of the page.

\section{Preliminary experiment}
\label{sec:preliminary}
\begin{figure}
    \centering
    \includegraphics[width=0.6\textwidth]{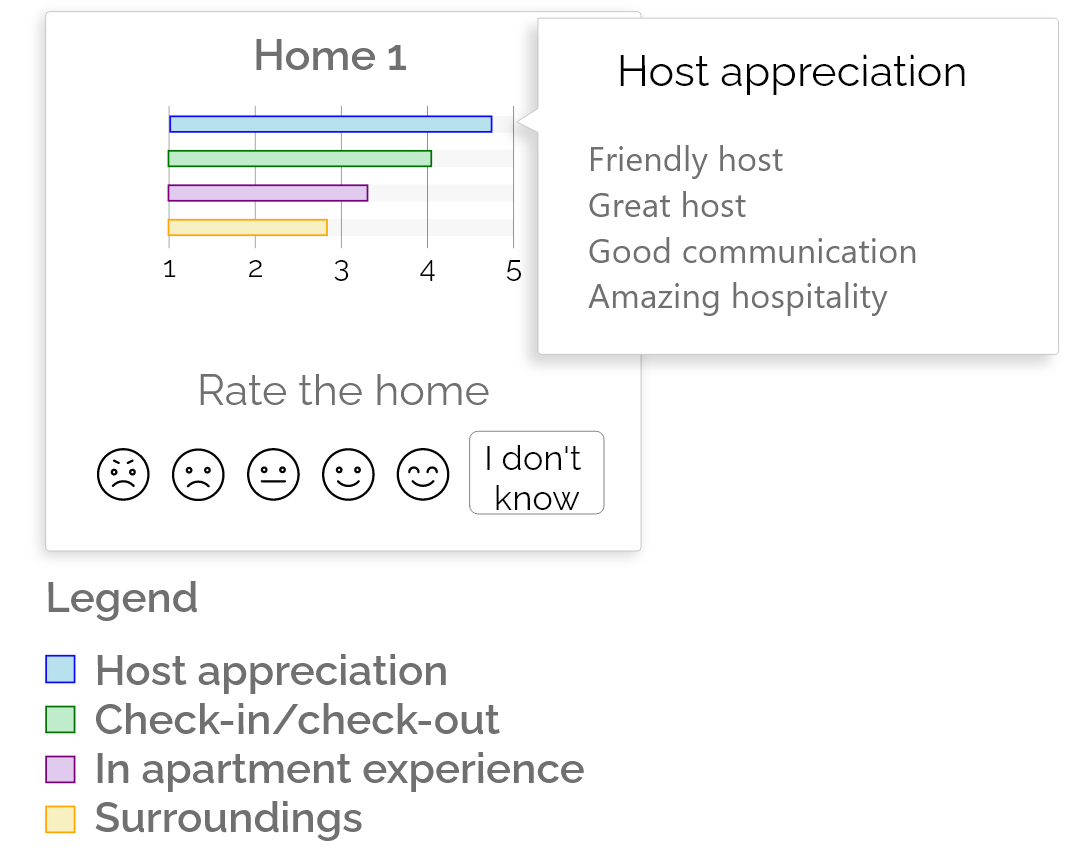}
    \caption{Portion of the mock-up user interface used in the preliminary experiment.}
    \label{fig:preliminary}
\end{figure}

We conducted a preliminary experiment to test a first mock-up of service-based justification model described in \citep{Mauro-etal:21c}; see Fig. \ref{fig:preliminary}. In the study we wanted to investigate the impact on decision making of the visualization of quantitative data that describes coarse-grained experience evaluation dimensions (bar graphs), combined with on-demand qualitative data concerning such dimensions (aspects manually crafted from item reviews). The model did not include the fine-grained evaluation dimensions of experience. In the following, we briefly describe this experiment.

11 participants, aged between 19 and 57, and having diverse background and familiarity with technology, joined in the study. 
They performed two tasks, each one requiring to evaluate 5 Airbnb homes presented by the mock-up:
\begin{itemize}
    \item 
    In the first task (\texttt{T1}), for each presented home the user could only view the bar graphs of the coarse-grained evaluation dimensions.
    \item 
    In the second one (\texttt{T2}), the user could also inspect the aspects and adjectives concerning the coarse-grained dimensions by clicking the respective bars.    
\end{itemize}  
 
In the post-task questionnaire of \texttt{T1}, 54.54\% of people declared that the given information was not enough to rate the homes. In \texttt{T2}, 27.27\% said that they would have preferred to receive more data about the homes while the other ones evaluated them.
This suggests that bar graphs alone are not sufficient for decision making because they do not enable the user to access any qualitative data to justify the values of the experience evaluation dimensions. When showing qualitative data the situation improves. However, more information about items is needed to help users make their selections.

The participants' comments provided insights about how they used the data provided by the mock-up. We report the two most interesting ones:
\begin{itemize}
    \item 
    In a real-word application, the bar graph is useful to filter out a home that does not deserve to be analyzed because it performs badly on an evaluation dimension that the user particularly cares about. For instance, a participant declared to have discarded a home having low \texttt{Host appreciation} to avoid interacting with difficult people.
    \item 
    Some people said that the qualitative data about evaluation dimensions (i.e., the aspects presented on-demand) are useful to implicitly ``tune" the values of the bars. For instance, suppose that a home $h$ receives a low evaluation regarding a dimension $d$, and that the justification of $d$ depends on aspects that are irrelevant to the user. Then, he or she might implicitly increase the evaluation of $h$. 
\end{itemize}

In summary, our preliminary experiment suggested that the bar graphs describing coarse-grained experience evaluation dimensions present relevant quantitative data about items but have to be coupled with qualitative data to help users make their decisions. This finding drove us to the development of the novel justification models proposed in the present paper.

\section{Study design}
\label{sec:study}
We carried out a user study to evaluate users' experience with the presented justification models and how the service-based ones perform with respect to Tintarev and Masthoff's explanatory aims of effectiveness (help users make good decisions), and satisfaction (increase the ease of use or enjoyment).

As observed by \cite{Tsai-Brusilovsky:21}, when users are free to explore a recommendation list, they tend to inspect both items placed at the top of the list and lower ranked ones. This means that the system should enable them to obtain an accurate impression of all the presented items, regardless of their suitability. To be sure that our models offer this type of support, similar to \citep{Tintarev-Masthoff:22}, in the study we showed both high and low-quality homes.

\subsection{Context}
We recruited adult participants by spreading an invitation message in public mailing lists and social networks. In that message, we specified that we had a preference for people who had previously used an online booking or e-commerce platform. Users joined the experiment voluntarily, without any compensation.

We carried out the user study by exploiting an interactive web-based test application linked in the invitation message. The application guided participants in all the steps of the study. 
To guarantee users' privacy, the application did not collect their names or any other identifying data. At the beginning of the interaction with the user, it generated a numerical identifier to tag the anonymous data it acquired during the interaction session.

We used a power analysis to figure out how many people we needed to obtain statistically significant experimental results. The following four parameters are used for this analysis:
\textit{Alpha ($\alpha$ = 0.05):} a $p$ value that indicates the probability threshold for rejecting the null hypothesis when there is no significant effect (Type I error rate).  
\textit{Power = 0.80:} the probability of accepting the alternative hypothesis if it is true (Type II error rate). 
\textit{Effect size = 0.35:} the expected effect size, i.e., the quantified magnitude of a result present in the population; our goal was to find medium-sized effects. 
\textit{Sample size N:} the required size of the sample of participants to maintain statistical power. The estimation of the sample size resulted in N = 55 that supports the actual statistical power of 80\%. We thus planned to collect at least 55 user tests.

\subsection{Method}
In the study, we applied a within-subjects approach. We managed each treatment condition as an independent variable and each participant received all the treatments. The test application counterbalanced the order of tasks to reduce the effects of practice and fatigue, as well as the impacts of result biases.
We did not impose any time limits to the execution of the test to let people free to explore the information provided by the application. Overall, the user study was organized as follows:

\begin{table*}[t]
\centering
\caption{Questionnaire about personal characteristics and mean values of the answers.}
\label{tab:pers_char}
\resizebox{1\textwidth}{!}{%
\begin{tabular}{llllllll}
\toprule
Construct & Factor & Statement & M(SD)\\ 
\midrule
\multirow[t]{4}{*}{\begin{tabular}[c]{l} \\ \\ \textit{Trust in} \\ \textit{Booking Systems} \\ (PC1)\end{tabular}}
& 1 & 
\begin{tabular}[c]{@{}l@{}}I tend to trust the suggestions generated by booking systems. \end{tabular} & 3.10(0.82) \\
& 2 & I think that the ratings given by other users are enough to book homes.& 3.19(0.86) \\
& 3 & 
\begin{tabular}[c]{@{}l@{}}I need to inspect the reviews given by other users to book homes.
 \end{tabular}&4.12(0.74)\\
& 4 & 
\begin{tabular}[c]{@{}l@{}}I need to inspect the description of the home to book it. \end{tabular}&4.31(0.70)\\
\midrule
\multirow[t]{4}{*}{\begin{tabular}[c]{l} \\ \textit{Trust in Technology}\\
(PC2)\end{tabular}}
& 1 & \begin{tabular}[c]{@{}l@{}}I feel technology never works. \end{tabular}& 1.66(0.58)\\
& 2 & 
\begin{tabular}[c]{@{}l@{}} I’m less confident in doing things when I use supporting technology.
 \end{tabular}&1.80(0.94)\\
& 3 & 
\begin{tabular}[c]{@{}l@{}}The usefulness of technology is highly overrated.\end{tabular}&1.92(0.86)\\
& 4 & 
\begin{tabular}[c]{@{}l@{}}I tend to trust a person/thing, even though I have little knowledge of it. \end{tabular}&2.83(0.89)\\

\bottomrule       
\end{tabular}
}
\end{table*}

\begin{enumerate}
    \item
    First, the test application prompted users to read the informed consent\footnote{The consent text can be found at this link: \url{https://bit.ly/3LypcZp}.} and declare that they were 18 years old or over. Moreover, it asked them to give their explicit agreement to participate in the study. Only those who positively answered both questions could take part in the experiment. 
    \item
    Next, the application asked users to fill in a questionnaire with their demographic information, cultural background, familiarity with booking and e-commerce platforms. The questionnaire is based on the ResQue recommender system questionnaire \citep{Pu-etal:11} and it is useful to understand the basic characteristics of the sample of users.
    The application also collected a set of constructs of Personal Characteristics (PC). We defined two constructs regarding the Trust in Booking Systems and the General Trust in Technology \citep{Tsai-Brusilovsky:21}; see Table \ref{tab:pers_char}. Participants answered the statements in the $\{$Strongly disagree, Disagree, Neither agree nor disagree, Agree, Strongly agree$\}$ scale, which we mapped to [1, 5]. The last column of the table shows the mean values of the participants' answers.
    
\begin{table*}[t]
\centering
\caption{Post-task questionnaire. Statements are grouped by user experience construct.}
\label{tab:post-taskQuestionnaire}
\resizebox{1\textwidth}{!}{%
\begin{tabular}{lllllll}
\toprule
Construct & Factor & Statement \\ 
\midrule
\multirow[t]{3}{*}{\begin{tabular}[c]{l}\\
 \textit{Perceived User Awareness}\\ 
\textit{Support} (U)\end{tabular}} 
& U1 & 
\begin{tabular}[c]{@{}l@{}}The information provided was sufficient for me to understand \\ what previous users think about the homes. \end{tabular}\\
& U3 & 
\begin{tabular}[c]{@{}l@{}}The information about the homes was easy to interpret and understand. \end{tabular}\\
& U4 & 
\begin{tabular}[c]{@{}l@{}}I quickly found the information about the homes.  \end{tabular}\\
\midrule
\multirow[t]{3}{*}{\begin{tabular}[c]{l} \textit{Interface Adequacy} (I)\end{tabular}}
& I1 & It was easy to understand why some homes were good and others not.\\
& I2 & 
\begin{tabular}[c]{@{}l@{}} I found the user interface very intuitive. \end{tabular}\\
& I3 & 
\begin{tabular}[c]{@{}l@{}}The user interface was sufficiently informative.\end{tabular}
\\
\midrule
\multirow[t]{3}{*}{\begin{tabular}[c]{l}
\textit{Satisfaction} (S)\end{tabular}} 
& S1 &  I think that I would like to frequently use this system to evaluate homes.
 \\
& S3 &  I thought this system to evaluate homes was easy to use.
\\
& S4 & I felt very confident using this system to evaluate homes.\\
\bottomrule       
\end{tabular}
}
\end{table*}

    \item
    Then, the application applied (in counterbalanced order) the justification models described in Sections \ref{sec:service-models} and \ref{sec:baselines} to interact with users.
    For each model, it asked participants to explore and rank five homes; then, it asked them to declare their level of agreement with the statements of the post-task questionnaire shown in Table \ref{tab:post-taskQuestionnaire}, which measures the user experience with the justification models. The statements are taken from \citep{Pu-etal:11,diSciascio-etal:19,Lewis:09} and capture users' perceptions of the user interface provided by each model. Participants answered in the $\{$Strongly disagree, \dots, Strongly agree$\}$ scale, mapped to [1, 5]. 
    Table \ref{tab:post-taskQuestionnaire} groups statements in three user experience constructs: \textit{Perceived User Awareness Support}, \textit{Interface Adequacy} and \textit{Satisfaction} that we use to gain a deeper view of the user experience with the justification models through Structural Equation Model analysis; see Section \ref{sec:STM}. 
\end{enumerate}
During the test, the application also administered the statements of the Curiosity and Exploration Inventory-II (CEI-II) questionnaire \citep{Kashdan:09} and of the Need for Cognition one \citep{Coelho-etal:20}. CEI-II allows to understand participants' motivation to seek out knowledge and new experiences (Stretching) and their willingness to embrace the novel, uncertain, and unpredictable nature of everyday life (Embracing). Need for Cognition investigates people’s tendency to engage in and enjoy thinking.

\section{Results}
\label{sec:experiments}
66 people joined in the user study from November 15 to December 15, 2021 but we excluded 7 of them because they did not pass the attention checks.
On average, the experiment lasted 35.45 minutes.

\subsection{User data}
\begin{itemize}
    \item 
    {\em Gender and age}. The 59 participants we considered for the test included 24 females, 33 males, 0 not-binary, 2 not declared, with the following age distribution: $\leq$ 20 (2 people), 21-30 (43), 31-40 (7),  41-50 (2), 51-60 (4) and  $>$ 60 (1 person). 
    \item 
     {\em Education level and background.} 13 subjects declared that they had attended the high school, 40 the university, and 6 stated that they had a PhD. 17 participants specified that they had a technical background, 31 a scientific one, 6 humanities and languages, 2 economics, and 4 other background. Furthermore, 46 people classified themselves as advanced computer users, 10 average ones, and 3 beginners.
     \item 
     {\em Familiarity with online booking or e-commerce platforms.} 18 people declared that they used those platforms a few times in a week, 26 a few times in a month, 14 a few times in a year and 1 person never used one before.
     \item 
     {\em Trust in Booking Systems (PC1).} 
     As shown in the last column of Table \ref{tab:pers_char}, participants moderately agreed with trusting the suggestions generated by booking systems (statement 1: M = 3.10, SD = 0.82). They concurred that, to book a home, they needed to inspect its reviews (statement 3: M = 4.12, SD = 0.74) and description (statement 4: M = 4.31, SD = 0.70). However, they only partially agreed that the ratings given by other users are enough to book homes (statement 2: M = 3.19, SD = 0.86). 
     \item
     {\em Trust in Technology (PC2).} 
     Participants positively evaluated technology. In fact, statements 1-3, which deny the trust in technology, have low mean values. Differently, participants tended to be suspicious towards people and things that they do not know (statement 4: M = 2.83, SD = 0.89).

\end{itemize}

\begin{table*}[t]
\centering
\caption{Post-task questionnaire results describing participants' experience with the justification models. Results are grouped by user experience construct. For each construct, three rows show the values obtained for the questions of Table \ref{tab:post-taskQuestionnaire} (factors). The ``Average" row reports the mean value of the factors. The highest values are in boldface. Stars denote the statistical significance of the difference between the best-performing model and the other ones. Significance levels: (***)$p < 0.01$, (**)$p < 0.05$, (*)$p < 0.1$.}
\label{tab:post-task}
\resizebox{1\textwidth}{!}{%
\begin{tabular}{lllllll}
\toprule
Construct                                                                                           & Factor & \multicolumn{5}{c}{Justification Model}                                                                                    \\ \cmidrule(l){3-7} 
	 &      & \modA          & \modB      & \modC              & \modelChen          & \modD           \\
	 &      & M(SD)            & M(SD)     & M(SD)            & M(SD)            & M(SD)            \\
	  \midrule
	  
	  \multirow[t]{4}{*}{\begin{tabular}[c]{l}\\  \textit{Perceived User Awareness}\\ \textit{Support}  \end{tabular}}   

& U1 & 3.61(0.95) & 3.44(1.10) & 2.59(1.12) & 3.08(1.16) & \textbf{3.73(1.08)} \\  
& U3 & \textbf{3.58(1.04)} & 3.42(1.04) & 3.56(1.10) & 3.07(1.19) & 3.07(1.22) \\  
& U4 & \textbf{3.80(1.16)} & 3.53(1.09) & 3.63(1.07) & 3.20(1.17) & 2.80(1.28) \\  

 & Average & \textbf{3.66(1.05)} & 3.46(1.07) & 3.26(1.19)** & 3.12(1.17)*** & 3.20(1.25)** \\
   \midrule
   
	  \multirow[t]{4}{*}{\begin{tabular}[c]{l} \textit{Interface Adequacy}\end{tabular}}  
& I1 & \textbf{3.46(1.06)} & 3.34(1.14) & 2.98(1.18) & 3.12(1.13) & 2.93(1.26) \\  
& I2 & 3.46(1.25) & 3.36(1.06) & \textbf{3.81(1.01)} & 3.44(1.10) & 3.53(1.02) \\  
& I3 & \textbf{3.64(0.98)} & 3.47(1.01) & 2.41(1.02) & 3.14(1.14) & 3.39(0.89) \\  

& Average & \textbf{3.52(1.10)} & 3.39(1.07) & 3.07(1.21)*** & 3.23(1.13)* & 3.28(1.09) \\
   \midrule
\multirow[t]{4}{*}{\begin{tabular}[c]{l}\\ \textit{Satisfaction} \end{tabular}}
	 
& S1 & \textbf{3.29(1.22)} & 3.10(1.18) & 2.58(1.10) & 2.86(1.17) & 3.00(1.08) \\  

& S3 & 3.61(0.98) & 3.41(0.89) & \textbf{3.85(0.96)} & 3.31(1.00) & 3.51(1.02) \\  
& S4 & \textbf{3.51(1.14)} & 3.37(0.91) & 2.90(1.21) & 3.05(1.09) & 3.27(1.05) \\ 

& Average & \textbf{3.47(1.12)} & 3.29(1.01) & 3.11(1.22)** & 3.07(1.10)** & 3.26(1.07) \\

	\bottomrule& & 
\end{tabular}
}
\end{table*}

\subsection{Analysis of participants' experience with the justification models}
Table \ref{tab:post-task} shows the results of the post-task questionnaire of Table \ref{tab:post-taskQuestionnaire}. 
According to a Kruskall-Wallis test, the differences between user experience constructs across the five models are statistically significant:
\begin{itemize}
    \item \textit{Perceived User Awareness Support}  [$H = 13.40$, $df = 4$, $p<0.008$];
    \item \textit{Interface Adequacy}  [$H = 10.21$, $df = 4$, $p<0.035$];
    \item \textit{Satisfaction}  [$H = 8.07$, $df = 4$, $p<0.084$].
\end{itemize}
Moreover, a \textit{post-hoc} comparison based on a Mann-Whitney test showed that:
\begin{itemize}
    \item 
    \textit{Perceived User Awareness Support}. \modA (M=3.66, SD=1.05) is the best justification model. The difference compared to \modC, \modelChen and \modD is statistically significant. Specifically, in \modA, participants perceived the information about the homes as the easiest to interpret and understand (statement U3). Moreover, \modA best supported them in quickly finding the data about homes (U4). On the other hand, \modD received the lowest evaluation regarding both the easy of interpretation/understanding of information about items, and the speed in finding it. As all the justification models present the amenities offered by the homes in the same way, we think that \modD challenged the participants because it does not summarize the reviews.
    Conversely, \modD best supported the understanding of previous guests's opinions about the homes (U1) because it presents the full set of reviews received by the visualized items. However, \modA is the second best model concerning this evaluation aspect.
    \item 
    \textit{Interface Adequacy}. Participants perceived \modA as the best justification model (M=3.52, SD=1.10); the differences with respect to \modC and \modelChen are statistically relevant. People felt that \modA helps the user understand why some homes are good or bad and is sufficiently informative. However, they perceived \modC as the most intuitive user interface, probably because it summarizes information in a simple text.
    \item 
    \textit{Satisfaction}. Participants perceived \modA (M=3.47, SD=1.12) as the best model and the differences with respect to \modC and \modelChen are statistically relevant. Users declared that they would like to frequently use \modA to evaluate homes.
    Moreover, they felt more confident using \modA than the other justification models. However, they perceived \modC as the easiest model to use. Also in this case, the reason might be the bare user interface of this model, which presents previous guests' experience with items using few sentences. Anyway, \modA is the second best model concerning the easy of use.
\end{itemize}

\begin{figure*}[t]
  \centering
  \includegraphics[width=1\textwidth]{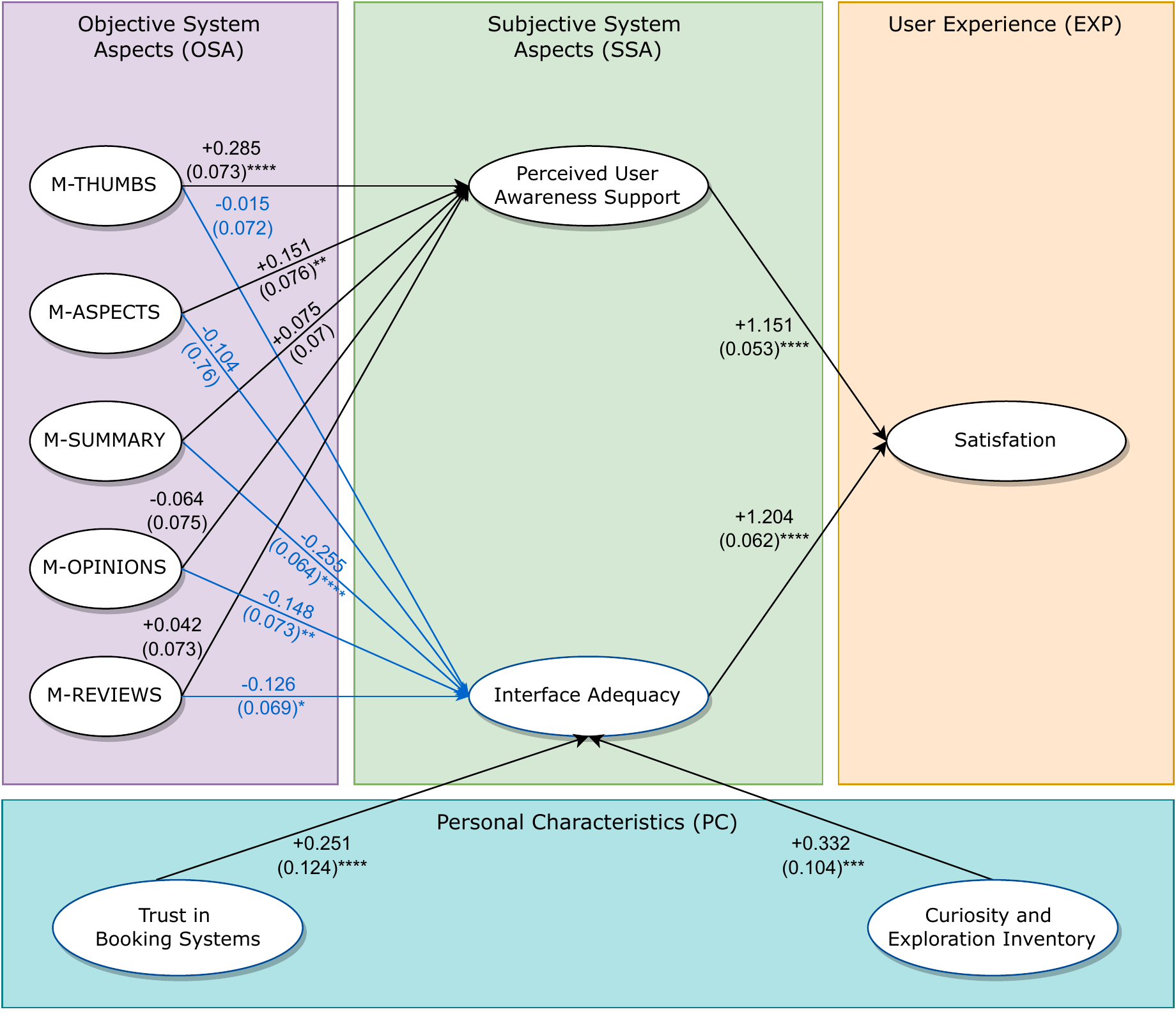}
  \caption{Structural Equation Model. 
  Significance levels:
  (****)$p < 0.001$, (***)$p < 0.01$, (**)$p < 0.05$, (*)$p < 0.1$. The numbers on the arrows represent the $\beta$-coefficients and standard error of the effect.}
  \label{fig:sem}
\end{figure*}

\subsection{Opting outs}
In the user study, the 59 participants evaluated 25 homes, thus providing 295 item ratings in total. Below, we report the number and rate of opting outs, which correspond to ``I don't know" selections and thus denote a lack of confidence in the evaluations:
\begin{itemize}
    \item \modA: 10 opting outs (3.39 \%);
    \item \modB: 15 (5.08 \%);
    \item \modC: 30 (10.17 \%);
    \item \modelChen: 33 (11.19 \%);
    \item \modD: 32 (10.85 \%).
\end{itemize}
These results are in line with the fact that \modA is the best justification model from the viewpoint of the \textit{Perceived User Awareness Support}. When using \modC and \modD, a definitely higher number of participants did not feel like evaluating the homes. This can be explained with the fact that (i) in spite of its simplicity, \modC poorly describes previous guests' opinions about the homes, and (ii) \modD forces the user to read the reviews to evaluate them. Differently, \modA shows a summary of consumer feedback but it also supports the on demand retrieval of detailed information about the homes.

\subsection{Structured Equation Model analysis}
\label{sec:STM}
We performed \cite{Knijnenburg:15}'s Structured Equation Model analysis to deeply understand the user experience with the justification models.
This analysis is useful to find the relationship between unobserved constructs (latent variables) by leveraging observable variables.

Based on the post-task questionnaire of Table \ref{tab:post-taskQuestionnaire}, we associated two constructs (\textit{Perceived User Awareness Support} and \textit{Interface Adequacy}) to the Subjective Systems Aspects (SSA) and one construct (\textit{Satisfaction}) to User Experience Aspects (EXP). We tested five justification models (Objective System Aspects) that we represented as dummy variables: \modA, \modB, \modC, \modD and \modelChen. Moreover, we selected the following constructs to represent the Personal Characteristics (PC): \textit{Trust in Booking Systems}, \textit{Trust in Technology},  \textit{Curiosity and Exploration Inventory} and \textit{Need for Cognition}. As these constructs include at least three statements each, they are good candidates for a Structured Equation Model.

We performed the Confirmatory Factor Analysis to examine the validity of the constructs. 
First, we computed the convergent validity to check that their statements are related. For this purpose, we examined the Average Variance Extracted ($AVE$) of each construct that must be over 0.50 to keep the validity. Then, we computed the discriminant validity to check if the statements of different constructs are too highly correlated.
To obtain the discriminant validity, the largest correlation value must be less than the squared root of the $AVE$ value of both factors. 
All the constructs respected the constraints:
\begin{itemize}
    \item \textit{Perceived User Awareness Support}: $AVE = 0.64$, $\sqrt{AVE(0.64)} = 0.80$, largest correlation = 0.59;
    \item \textit{Interface Adequacy}: $AVE = 0.47$, $\sqrt{AVE(0.47)} = 0.69$, largest correlation = 0.68;
    \item \textit{Satisfaction}: $AVE = 0.62$, $\sqrt{AVE(0.62)} = 0.79$, largest correlation = 0.70.
\end{itemize} 

Fig. \ref{fig:sem} shows the Structural Equation Model we obtained.
The graph reports the dependencies, $\beta$-coefficients and standard error values that indicate the correlations between the constructs.
We removed \textit{Trust in Technology} and \textit{Need for Cognition} from the Personal Characteristics because during the iterations they had high $p$ values.

All the models have a positive effect on the \textit{Perceived User Awareness Support} except \modelChen that has a neutral correlation. Specifically, \modA has the strongest positive correlation (+0.285; $p>0.001$) followed by \modB (+0.151; $p>0.05$). This is in line with the results of the post-task questionnaire, in which participants perceived \modA and \modB as the best models from this viewpoint. Moreover, since there is a positive correlation between the \textit{Perceived User Awareness Support} and the \textit{Satisfaction} construct (+1.151; $p<0.001$), we infer that \modA also satisfies users in the analysis of previous guests' feedback when they explore the homes.

All the models negatively influence the \textit{Interface Adequacy}, probably because their user interfaces require some effort to understand the representation of consumer feedback. However, \modA has the less negative value, i.e., it is better than the other models.
Notice that the \textit{Interface Adequacy} and \textit{Satisfaction} constructs are positively correlated (+1.204; $p<0.001$). From this information we infer that the models that best correlate with the \textit{Interface Adequacy} also lead to higher satisfaction. See Table \ref{tab:post-task}.

We notice a positive correlation between the \textit{Curiosity and Exploration Inventory} construct and the \textit{Interface Adequacy} one with lower statistical significance (+0.332; $p<0.05$).
Finally, there is a positive correlation between the \textit{Trust in Booking Systems} construct and the \textit{Interface Adequacy} one  (+0.251; $p<0.001$). This finding suggests that the people who have high trust in booking systems perceive the user interfaces of the justification models positively and they are more satisfied than the other users.

\begin{table}[]
\centering
\caption{Post-task questionnaire results grouped by CEI-II value. The highest values for each group of participants are in boldface. Stars denote the statistical significance of the difference between the best-performing model and the other ones. Significance levels: (***)$p < 0.01$, (**)$p < 0.05$, (*)$p < 0.1$.}
\label{tab:cei}
\resizebox{\textwidth}{!}{%
\begin{tabular}{llll}
	\toprule
	\multicolumn{1}{l}{}  
	&  & CEI-II$<$3.5 & CEI-II$>=$3.5 \\
	\midrule
	\multicolumn{1}{l}{\multirow{3}{*}{\modA}} & \textit{Perceived User Awareness Support} & \textbf{3.67}  & \textbf{3.66}  \\
	\multicolumn{1}{l}{}  &
	\textit{Interface Adequacy} & \textbf{3.64}  & 3.44 \\
	\multicolumn{1}{l}{}  & 
	\textit{Satisfaction} & \textbf{3.51}  & \textbf{3.44}  \\
	\midrule
	\multirow{3}{*}{\modB} & 
	\textit{Perceived User Awareness Support} & 3.29 & 3.58  \\
	& \textit{Interface Adequacy} & 3.29 & \textbf{3.46}\\
    & \textit{Satisfaction} & 3.21 & 3.35  \\
	\midrule
	\multirow{3}{*}{\modC}  & 
	\textit{Perceived User Awareness Support} & 3.11**  & 3.36 \\
    & \textit{Interface Adequacy} & 3.08** & 3.06**\\
    & \textit{Satisfaction}  & 2.94** & 3.22 \\
	\midrule
	\multirow{3}{*}{\modelChen}  & 
	\textit{Perceived User Awareness Support} & 2.92***  & 3.26* \\
    & \textit{Interface Adequacy} & 3.06** & 3.35\\
    & \textit{Satisfaction} & 2.92** & 3.18 \\
	\midrule
	\multirow{3}{*}{\modD}  & 
	\textit{Perceived User Awareness Support} & 3.24  & 3.17 \\
    & \textit{Interface Adequacy} & 3.26 & 3.30\\
    & \textit{Satisfaction} & 3.13 & 3.35** \\
	\bottomrule
\end{tabular}%
}
\end{table}

\subsection{User experience analysis based on personality traits and cognitive styles}
To further understand participants' perceptions of the justification models we analyzed the {\em Perceived User Awareness Support}, {\em Interface Adequacy} and {\em Satisfaction} by taking personality traits and cognitive style into account. 

\subsubsection{Curiosity trait}
We grouped participants depending on the value they obtained when they answered the Curiosity and Exploration Inventory-II (CEI-II) questionnaire, which measures the motivation to seek out knowledge and new experiences (Stretching) and the willingness to embrace the novel, uncertain, and unpredictable nature of everyday life (Embracing). Participants with a high (or low) CEI-II value have high (low) Stretching and Embracing levels. The first group includes 24 people having a CEI-II value $<$3.5; the second one includes 35 people with a value $\geq$3.5. We could not obtain more balanced groups according to the results of the questionnaire. 
Table \ref{tab:cei} shows the user experience results:
\begin{itemize}
    \item 
    The participants having low CEI-II value evaluated \modA as the best performing model in all the user experience constructs; see the values in bold in column ``CEI-II$<$3.5". The difference between this model and \modC and \modelChen is statistically significant. The second best model is \modB, regarding all three constructs.
    \item 
    The situation for the participants having a high CEI-II value is mixed but supports the service-based justification models we propose. In this case, \modA obtained higher values than the baselines in all the constructs. However, \modB excelled in \textit{Interface Adequacy}.
\end{itemize}
These findings confirm that \modA is the best justification model, regardless of the user's curiosity. However, the \textit{Interface Adequacy} results deserve further investigation. 
The main difference between \modA and \modB is the fact that \modA summarizes consumer feedback through bar graphs and thumbs up/down. 
Differently, \modB combines bar graphs with interactive filters representing the individual adjectives that are associated with the aspects. 
We can thus explain the diverse participants' perceptions with the fact that, as users having a low CEI-II value don't like to seek out knowledge, they prefer a model that offers a simple and quick (schematic) summary of consumer feedback, rather than a user interface offering advanced data exploration functions. Conversely, highly curious people are inclined towards interacting with fine-grained widgets and a larger amount of information.

\begin{table}[]
\centering
\caption{Post-task questionnaire results grouped by Need for Cognition (NfC) value. We use the same notation as in Table \ref{tab:cei}.}
\label{tab:need}
\resizebox{\textwidth}{!}{%
\begin{tabular}{llll}
	\toprule
	\multicolumn{1}{l}{} & & NfC$<$3.5 & NfC$>=$3.5 \\
	\midrule
	\multicolumn{1}{l}{\multirow{3}{*}{\modA}} &
	\textit{Perceived User Awareness Support} & \textbf{3.73} & \textbf{3.61} \\
	\multicolumn{1}{l}{}  & 
	\textit{Interface Adequacy} & \textbf{3.68} & 3.39 \\
	\multicolumn{1}{l}{} & 
	\textit{Satisfaction}  & \textbf{3.60} & 3.36 \\
	\midrule
	\multirow{3}{*}{\modB} & 
	\textit{Perceived User Awareness Support} & 3.63 & 3.33 \\
	& \textit{Interface Adequacy} & 3.51 & 3.29 \\
    & \textit{Satisfaction} & 3.44 & 3.18 \\
	\midrule
	\multirow{3}{*}{\modC}                     & \textit{Perceived User Awareness Support} & 3.19** & 3.31 \\
    & \textit{Interface Adequacy}  & 2.99*** & 3.13* \\
	& \textit{Satisfaction} & 3.00** & 3.19 \\
	\midrule
	\multirow{3}{*}{\modelChen}  & 
	\textit{Perceived User Awareness Support} & 3.21** & 3.05*** \\
	 & \textit{Interface Adequacy} & 3.36 & 3.13 \\
	  & \textit{Satisfaction} & 3.04** & 3.10 \\
	\midrule
	\multirow{3}{*}{\modD} & 
	\textit{Perceived User Awareness Support} & 3.03** & 3.33 \\
	& \textit{Interface Adequacy} & 3.08** & \textbf{3.44} \\
    & \textit{Satisfaction} & 3.05** & \textbf{3.42} \\
	\bottomrule
\end{tabular}%
}
\end{table}

\subsubsection{Need for Cognition}
We also grouped participants depending on the value they obtained in the Need for Cognition (NfC) questionnaire, which measures the tendency to engage in and enjoy thinking. The first group includes the participants with a NfC value $<$3.5 (26 people); the second includes those having a NfC value $\geq$3.5 (33 people).
Table \ref{tab:need} shows the results of the user experience analysis:
\begin{itemize}
    \item
    On the group of participants having NfC$<$3.5, \modA is the best performing model and \modB is the second best. Both models obtained higher user experience values than the baselines in all the constructs and most differences between \modA and the baselines are statistically significant.
    \item 
    On the people having a high NfC value, \modA obtained the best evaluation regarding the \textit{Perceived User Awareness Support} construct, followed by \modB and \modD. However, \modD obtained the best results concerning \textit{Interface Adequacy} and \textit{Satisfaction}.
    \end{itemize}
In summary, both participant groups perceived that our service-based justification models support awareness in a more efficacious way than the baselines. Moreover, \modA emerged as the best model for the people with low NfC while \modD was the best one for the other users.
We explain these observations with the fact that the people with low Need for Cognition prefer the models offering a schematic and organized summary of consumer feedback, such as bar graphs and thumbs. The other models are more suitable to the users having a high Need for Cognition because these people prefer to autonomously analyze and interpret data, as it happens in \modD.

\subsubsection{Discussion}
\modA offers a concise view of information that requires little effort to understand consumer feedback. This makes it particularly good for the people with low curiosity or need for cognition. Moreover, \modA enriches the summary of information by enabling people to inspect the details of items they care about. For this reason, all users, regardless of their CEI-II and NfC values, appreciate its awareness support.

Differently, the people with high Need for Cognition prefer models such as \modD and \modelChen (which allow direct access to information without summarization) as far as the interface adequacy and satisfaction are concerned. This is probably due to the fact that these models are less assistive and offer larger freedom in data exploration. For instance, they show the plain text of the item reviews or a detailed list of features to analyze. In all such cases, users actively interact with the system to extract the data they are interested in.

\begin{table}[t]
\centering
\caption{Log analysis. The Total column reports mean values for all the participants of the user study. The last two columns refer to the CEI-II groups.}
\label{tab:log-cei}
\resizebox{\textwidth}{!}{%
\begin{tabular}{@{}lllllll@{}}
\toprule
\multicolumn{1}{l}{}
& & Total  & CEI-II$<$3.5  & CEI-II$>=$3.5 \\ 
\midrule
\multirow{3}{*}{\modA}    
& Time spent to explore 5 homes & 175.58 &205.79 & 154.86\\
& \#clicks on the bar graphs & 38.17  & 36.83 & 39.09\\
& \#clicks on fine-grained dimensions & 15.41 &	18.83 & 13.06\\
& \#clicks to view more aspects & 14.36 &	17.13 &	12.46 \\
& \#clicks on thumbs up/down & 24.20  & 27.08  & 22.23\\ 
\midrule
\multirow{3}{*}{\modB} 
& Time spent to explore 5 homes & 169.03 & 193.92 & 151.97 \\
& \#clicks on the bar graphs & 29.00  & 36.25  & 24.03\\
& \#clicks on fine-grained dimensions & 13.24 &	17.29	& 10.46\\
& \#clicks to view more aspects & 12.36 &	15.92 &	9.91 \\
& \#clicks on the aspects & 24.56  & 26.88  & 22.97 \\ 
\midrule
\modC 
& Time spent to explore 5 homes & 76.14 & 89.08  & 67.26 \\ 
\midrule
\multirow{3}{*}{\modelChen} 
& Time spent to explore 5 homes & 152.54 & 185.88 & 129.69\\
& \#clicks on aspects & 59.61 & 64.08  & 56.54 \\
& \#visualized aspects & 80.66  & 88.25  & 75.46 \\ 
\midrule
\multirow{2}{*}{\modD}    
& Time spent to explore 5 homes & 270.07 & 310.71 & 242.20\\
& \#visualized reviews & 30.24  & 32.29  & 28.83 \\ 
\bottomrule
\end{tabular}%
}
\end{table}

\subsection{Log analysis}
Tables \ref{tab:log-cei} and \ref{tab:log-nfc} summarize the analysis of the logged actions considering both the complete group of participants and the splits by CEI-II or NfC value. 
For each justification model, the tables report the mean time spent to explore the five homes presented during the user study, and the mean number of clicks or visualized data during the interaction with the test application. Specifically:
\begin{itemize}
    \item 
    ``\#clicks on the bar graphs" (\modA, \modB) is the mean number of clicks to open the widget that shows the fine-grained dimensions associated with a specific coarse-grained one.
    \item
    ``\#clicks on fine-grained dimensions" (\modA, \modB) is the mean number of clicks to view the aspects of a fine-grained dimension.
    \item 
    ``\#clicks to view more aspects" (\modA, \modB) is the mean number of clicks to visualize more aspects of a fine-grained dimension.
    \item 
    ``\#clicks on thumbs up/down" (\modA) is the mean number of clicks to view the positive/negative quotes of reviews concerning an aspect.
    \item 
    ``\#clicks on aspects" (\modB) is the mean number of clicks to view the quotes of the reviews mentioning a specific aspect.
    \item 
    ``\#visualized aspects" (\modelChen) is the mean number of displayed aspects, given that the user can scroll down the list.
    \item
    ``\#visualized reviews" (\modD) is the mean number of visualized reviews, given that the user can scroll down the list.
\end{itemize}

\begin{table}[t]
\centering
\caption{Log analysis. We use the same indicators and notation as in Table \ref{tab:log-cei} and we repeat column Total for clarity. Participants are grouped by NfC.}
\label{tab:log-nfc}
\resizebox{\textwidth}{!}{%
\begin{tabular}{@{}lllllll@{}}
\toprule
\multicolumn{1}{l}{}  &  & Total  & NfC$<$3.5 & NfC$>=$3.5\\ 
\midrule
\multirow{3}{*}{\modA}    
& Time spent to explore 5 homes & 175.58 & 124.96  & 215.45 \\
& \#clicks on the bar graphs & 38.17 & 25.73 & 47.97\\
& \#clicks on fine-grained dimensions & 15.41 &	11.46 & 18.52\\
& \#clicks to view more aspects & 14.36 & 10.96 & 17.03
\\
& \#clicks on thumbs up/down & 24.20 & 15.81 & 30.82\\ 
\midrule
\multirow{3}{*}{\modB}    
& Time spent to explore 5 homes & 169.03 & 139.73  & 192.12 \\
& \#clicks on the bar graphs & 29.00  & 22.23 & 34.33\\
& \#clicks on fine-grained dimensions & 13.24 &	12.23	& 14.03\\
& \#clicks to view more aspects & 12.36 &	11.54&	13.00 \\
& \#clicks on aspects & 24.56  & 19.15 & 28.82\\ 
\midrule
\modC 
& Time spent to explore 5 homes  & 76.14 & 69.92 & 81.03\\ 
\midrule
\multirow{3}{*}{\modelChen} 
& Time spent to explore 5 homes & 152.54 & 104.58 & 190.33 \\
& \#clicks on the aspects & 59.61  & 46.85 & 69.67\\
& \#visualized aspects & 80.66 & 72.46 & 87.12\\ 
\midrule
\multirow{2}{*}{\modD}    
& Time spent to explore 5 homes & 270.07 & 225.04 & 305.55\\
& \#visualized reviews & 30.24  & 28.81 & 31.36\\ 
\bottomrule
\end{tabular}%
}
\label{tab:log-nfc}
\end{table}

\subsubsection{Complete group of participants}
\modD engaged participants in the interaction for the longest time because, to make an opinion about the homes, they had to read and analyze the reviews. In average, people visualized about 30 reviews, i.e., 6 for each home. As the user interface typically shows a maximum of 3 reviews (unless they are very short), we infer that participants scrolled down the review lists to inspect more opinions about the homes. 

The opposite situation holds for \modC that summarizes the opinions emerging from the reviews in a short text.

\modA and \modB are in the middle of the ranking of all the justification models and engaged participants a bit longer than \modelChen. Moreover, people spent slightly more time in interacting with \modA than \modB. We explain these findings by analyzing the clicks on the widgets:
\begin{itemize}
    \item 
    When participants used \modA, on average they expanded the bar graphs 38.17 times, which roughly corresponds to 8 times per home. This means that they tended to go back and forth from one to the other one. They opened the widgets of specific fine-grained dimensions (mean total number: 15.41 times) and expanded the evaluation dimensions (14.36) to see the complete list of aspects about 3 times per home. The mean number of clicks on the thumbs up/down to view the quotes from the reviews is 24.20, i.e., about 5 clicks per home. This finding suggests that people considered the thumbs (which also show the number of reviews supporting the evaluations) a good synthesis of previous guests' perceptions of the homes. Thus, they did not need to inspect many quotes from the reviews.
    \item 
    On \modB, participants explored the coarse-grained evaluation dimensions about 6 times per home (mean total number: 29). Moreover, they performed about the same number of clicks on the fine-grained dimensions and they clicked about 5 aspects per home (24.56), similar to the number of clicks on thumbs up/down of \modA. As each aspect might have more than one adjective, this finding suggests that participants selectively inspected the quotes associated with the adjectives.
    \item
    On \modelChen, participants visualized several aspects of the presented homes (about 16 per home, 80.66 in total).
    The reason is that \modelChen directly shows the list of aspects, without grouping them by fine-grained dimension. Therefore, to make an opinion about a home, users had to check several aspects, looking for the relevant ones. 
\end{itemize}

\subsubsection{Curiosity trait}
Table \ref{tab:log-cei} shows the log analysis on the groups of participants having low and high CEI-II value respectively. The time spent to evaluate the 5 homes of the experiment, and the clicking behavior on the user interfaces, have similar trends to those of the complete user group, confirming the interpretation we gave in the previous section. 
However, when comparing behavior of the participants with low and high CEI-II values on the same models, we can see that the people with low CEI-II spent more time in the evaluation than the other group. Moreover, they performed a larger number of clicks to inspect the aspects of the homes or their reviews. 
This is probably due to the fact that the less curious users are less efficient in finding the information they need and thus browse the user interface longer to make an opinion about the homes. 

\subsubsection{Need for Cognition}
Table \ref{tab:log-nfc} shows the log analysis on the groups of participants organized by NfC value.
Also in this case, \modD engages participants in the longest interaction to evaluate homes and \modC has the shortest evaluation times. However, there are differences concerning \modA and \modB. The people having high NfC exhibited a similar pattern in the evaluation of homes as the complete group of participants. Differently, when using \modA, low NfC users spent less time in the evaluation but performed a higher number of clicks than in \modB. While this finding seems to be contradictory, we explain it with the fact that, by summarizing the evaluation of aspects through thumbs up/down, \modA supports a more efficient evaluation of homes than \modB that requires to investigate individual aspects.

\subsubsection{Discussion}
The log analysis, either performed on the whole group of participants, or on the subgroups split by curiosity trait/cognitive style, shows that \modD engages users for the longest time to evaluate homes. This is because, other than checking the mean rating received by the items, it requires reading possibly long lists of reviews. In contrast, \modC engages users for the shortest time because it proposes a summary of consumer feedback.

The incremental access to data supported by \modA and \modB through bar graphs and fine-grained information exploration widgets (thumbs and clickable adjectives) engages users in the interaction a bit longer than \modC and \modelChen. However, the analysis of the clicks shows that participants evaluated the homes by inspecting a relatively low amount of data. Moreover, \modA and \modB received the best evaluation of their information awareness support from all the participants. Thus, the longer time spent in interacting with the widgets can be interpreted as a sign of interest.

\section{Lessons learned}
\label{sec:discussion}
The main finding that emerged from our study is that the \textit{Perceived User Awareness Support} of our service-based justification models (especially by \modA) is higher than that of the state-of-the art models we considered. As this support is key to decision making and it is agnostic to the algorithms underlying item suggestion, our experiments encourage the introduction of service-based justification in recommender systems.
However, participants' perception of the interface adequacy and satisfaction with these models depends on their curiosity traits and cognitive styles: 
\begin{itemize}
\item
The analysis of the user feedback shows that the overall group of participants evaluated \modA and \modB as the best justification models on all the user experience constructs. Moreover, the Structured Equation Model analysis, and the fact that the service-based models received the smallest numbers of opting outs in the evaluation of the homes, confirms their superiority to the baselines. From the log analysis we also found that the service-based justification models enabled participants to make an opinion about the homes by inspecting a relatively low amount of information. 
\item
However, the highly curious participants perceived the \textit{Interface Adequacy} of \modB (that supports the inspection of the individual adjectives of aspects) higher than that of \modA (which summarizes opinions by means of thumbs up/down). 
Moreover, participants having low Need for Cognition preferred \modD, which directly presents the item reviews, as far as the \textit{Interface Adequacy} and \textit{Satisfaction} constructs are concerned.
\end{itemize}
We interpret these findings by saying that the less curious users, and those with a low Need for Cognition, are comfortable with a justification model that organizes and summarizes data around the stages of interaction with items, offering a quick way to retrieve the data they care about. Differently, curious people and users with a high Need for Cognition benefit from the service-based organization of information but prefer to be autonomous in the analysis of consumer feedback.
In line with the findings of other works, such as \citep{Millecamp-etal:22,Kouki-etal:19,Kouki-etal:20}, this suggests that, to support all users in an informed item selection, we should extend service-based justification of results with the personalization of the user interface to the user's characteristics.

\section{Limitations and future work}
\label{sec:limitations}
As our justification models are broadly applicable but they fail to fully satisfy the users having a high Need for Cognition, we plan to personalize their user interfaces, e.g., by enabling the user to dynamically select the justification model to be used on an individual basis. 
Another possibility is to combine the information exploration functions offered by different models in a way that enables the user to choose the preferred ones. For instance, the service-based justification model might also offer a widget that enables the user to make a direct access to item reviews. 
In our future work, we plan to improve the user interface of our models by integrating, possibly in a personalized way, some of the functions offered by the other models. For this purpose, it will be crucial to understand how to recognize the user's Need for Cognition value, or how to make the user interface adaptable so that the user can select the preferred interaction features. The adaptability of the user interface might also be strategic to comply with user interface requirements implied by the level of risk of the task that users have to perform. In the user study, participants engaged in a realistic but artificial task because they did not spend their money to book the homes. Therefore, we could not measure their preferences for the user interfaces of the justification models under this perspective.

Another limitation of our service-based models is that they sort item aspects on a relevance basis by considering the number of reviews that mention them. In other words, we currently adhere to a conformity principle to decide which aspects are worth promoting. While the opinion of the mass is important, the user's interests might be considered as well to reflect individual priorities, as suggested in works such as \citep{Pu-Chen:07} and since the early faceted information exploration approaches \citep{Tvarozek-etal:08,Abel-etal:11}.
In our future work, we plan to tailor the presentation of aspects by analyzing the user's preferences for the fine-grained evaluation dimensions corresponding to the stages of interaction with items, and by steering the content presented in the user interface accordingly. This type of personalization might be applied to both aspect-based and service-based recommender systems, which take item aspects into account.

In our future work, we also plan to extend our Service Blueprint to model amendment episodes (e.g., due to exceptions at the consumers' side, or failures in the appliances of the rented home) that might involve interacting with the host, and thus impact consumers' experience as well.

Finally, we plan to test our service-based justification models on other domains such as the e-commerce one to assess their applicability to heterogeneous types of items. In this respect, it is worth mentioning that the specification of a new application domain is frequently supported by existing service models that can be adapted to the peculiarities of the selected domain. For instance, there are Service Blueprints defining the sales of products in online retailer platforms \citep{Gibbons:17} and food and beverage service systems \citep{Nam-etal:18}. Specifically, in \cite{Gibbons:17}'s blueprint, customer actions include visiting the website, visiting the store and browsing for products, discussing options and features with a sales assistant, purchasing the product, getting a delivery-date notification, and finally receiving the product. Moreover, the onstage/visible contact employee actions include the event of meeting customers in the store, the interaction with a chat assistant on the website to get more information about products, and so forth. Finally, the physical evidence layer includes, among the others, the products, the physical stores, the website, and the tutorial videos available on the website. All these elements of the blueprint can be exploited to define coarse-grained and fine-grained experience evaluation dimensions for our models.

\section{Ethical issues}
\label{sec:ethics}
In planning the user study we complied with literature guidelines on controlled experiments\footnote{\url{https://www.tech.cam.ac.uk/research-ethics/school-technology-research-ethics-guidance/controlled-experiments}} \citep{Kirk:13}.  
Through the user interface of our test application, and the informed consent, we notified participants about their rights:
(i) the right to stop participating in the experiment at any time, without giving a reason;
(ii) the right to obtain further information about the purpose, and the outcomes of the experiment; 
(iii) the right to have their data anonymized. 

We did not collect participants' names, nor any data that could be used to identify them. During the user study, and the analysis of its results, we worked with anonymous codes, generated on the fly when users started the experiment. 
As described in Section \ref{sec:study}, before starting the experiment, participants had to (i) confirm that they were 18 years old or over, (ii) read a consent form describing the nature of the experiment and their rights (\url{https://bit.ly/3LypcZp}), and (iii) confirm that they had read and understood their rights by clicking on the user interface of the test application. 
Every participant was given the same instructions before the experimental tasks. Our experiment has been approved by the ethics committee of the University of Torino (Protocol Number: 0421424).

\section{Conclusions}
\label{sec:conclusions}
We described a novel approach to justify recommender systems results based on an explicit representation of the service underlying the suggested items. Our goal was that of enhancing users' awareness about the suggested items with a holistic view of previous consumers' experience that considers the services and actors that the user might encounter during the overall interaction with an item, from its selection to its usage.

The existing work on the justification of recommender systems' results generates short descriptions \citep{Musto-etal:19b, Musto-etal:20} or flat lists of aspects \citep{Chen-etal:14} to present the suggested items. Differently, we leverage the Service Blueprints to summarize consumer feedback at different granularity levels, taking the stages of the service underlying items into account. Our approach supports an incremental access to the information about previous consumers' experience with items. By explicitly representing service-based data, the system provides users with a holistic description of items aimed at enhancing their awareness of the available options and their confidence in item evaluation.

We applied our approach to the home booking domain that is is particularly challenging because it exposes the user to the interaction with multiple entities, such as homes, hosts and surroundings, which might dramatically impact the renting experience.
A user test involving 59 participants has shown that our service-based justification models obtain higher results than state-of-the-art baselines as far as the \textit{Perceived User Awareness Support}, \textit{Interface Adequacy} and \textit{Satisfaction} user experience constructs are concerned. However, we noticed that the people having a high Need for Cognition prefer a direct analysis of consumer feedback, without the mediation of a summarization model. 
These results encourage the adoption of service-based justification models in recommender systems but they suggest to investigate the adaptation of these models to the individual user.

\bmhead{Acknowledgments}
This work was funded by the University of Torino. We are grateful to Gianmarco Izzi and Livio Scarpinati for their contribution to the software development of the test application we used. We also thank the anonymous reviewers who helped us improve the paper with their comments.

\begin{appendices}

\section{Rotated figures}\label{secA1}

 \begin{sidewaysfigure}
    \centering
    \includegraphics[width=1\columnwidth]{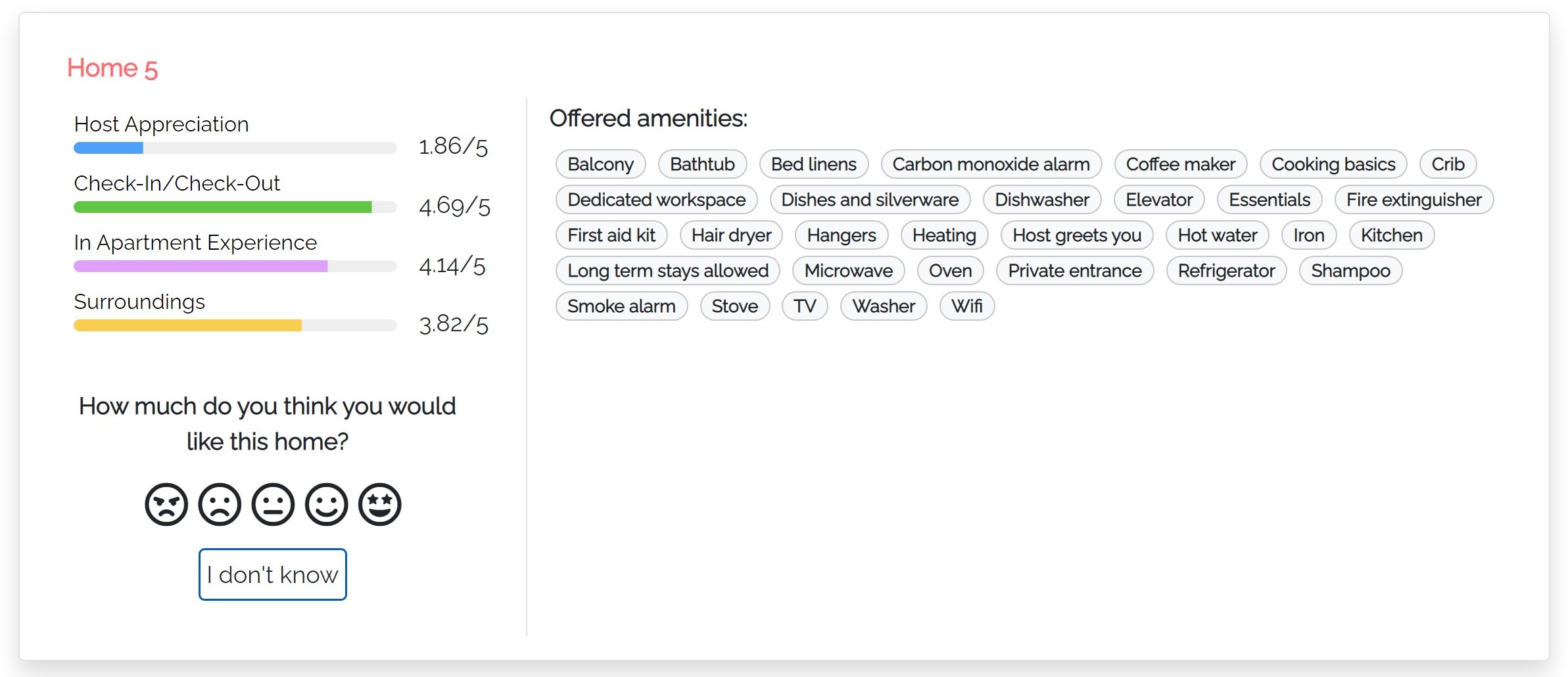}
    \caption{Portion of the user interface shared by the proposed justification models.}
    \label{fig:ui-appendix}
\end{sidewaysfigure}

\begin{sidewaysfigure}
    \centering
    \includegraphics[width=1\columnwidth]{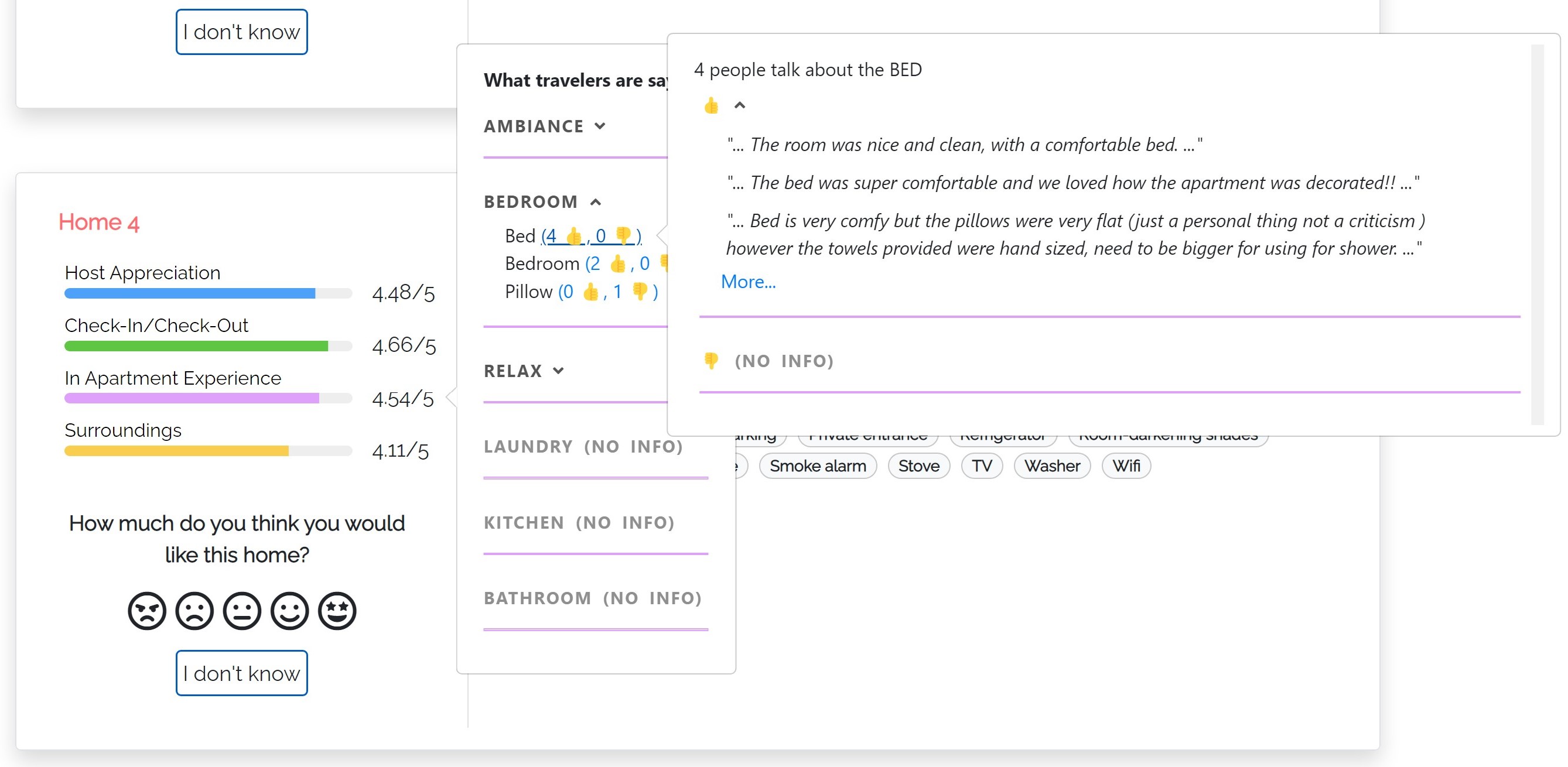}
    \caption{User interface of justification model \modA.}
    \label{fig:modelA-appendix}
\end{sidewaysfigure}

\begin{sidewaysfigure}
    \centering
    \includegraphics[width=1\columnwidth]{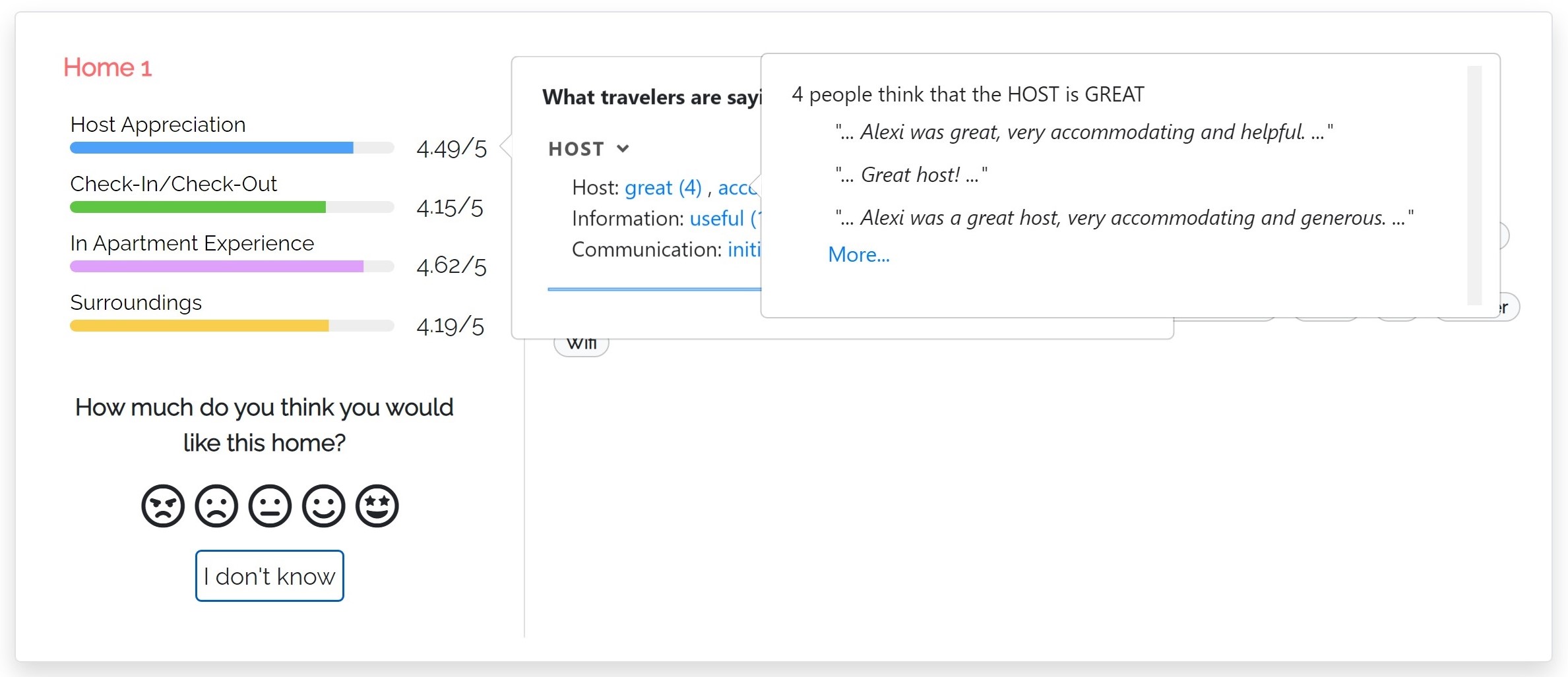}
    \caption{User interface of justification model \modB.}
    \label{fig:modelB-appendix}
\end{sidewaysfigure}

\begin{sidewaysfigure}
    \centering
    \includegraphics[width=1\columnwidth]{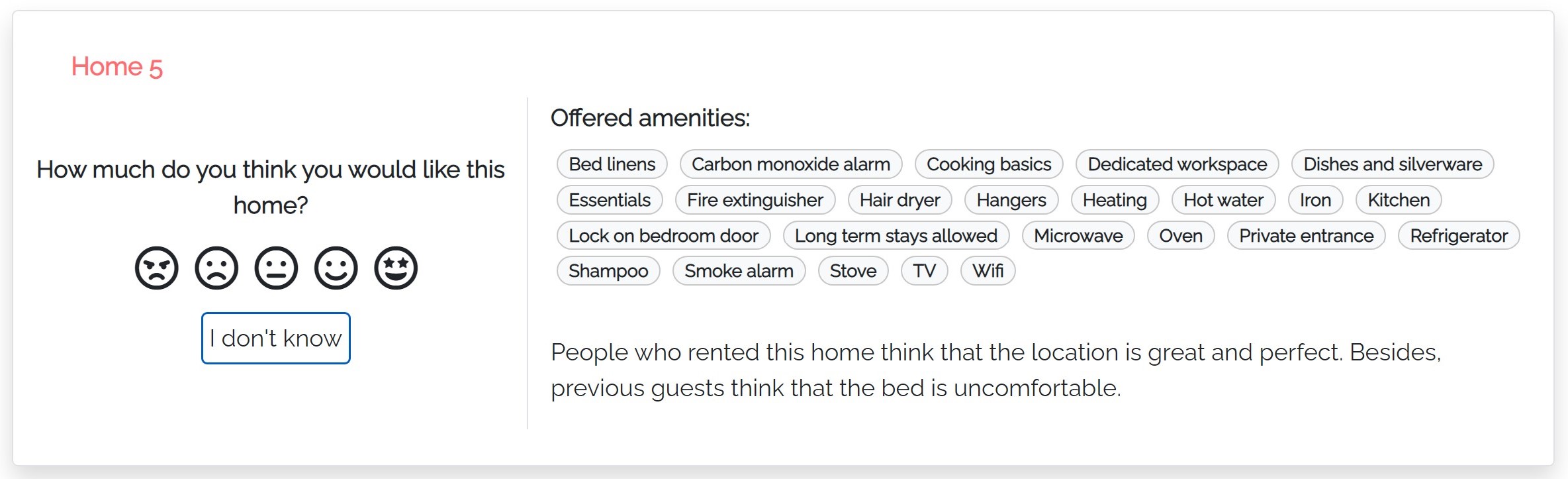}
    \caption{User interface of justification model \modC.}
    \label{fig:modelC-appendix}
\end{sidewaysfigure}

\begin{sidewaysfigure}
    \centering
    \includegraphics[width=1\columnwidth]{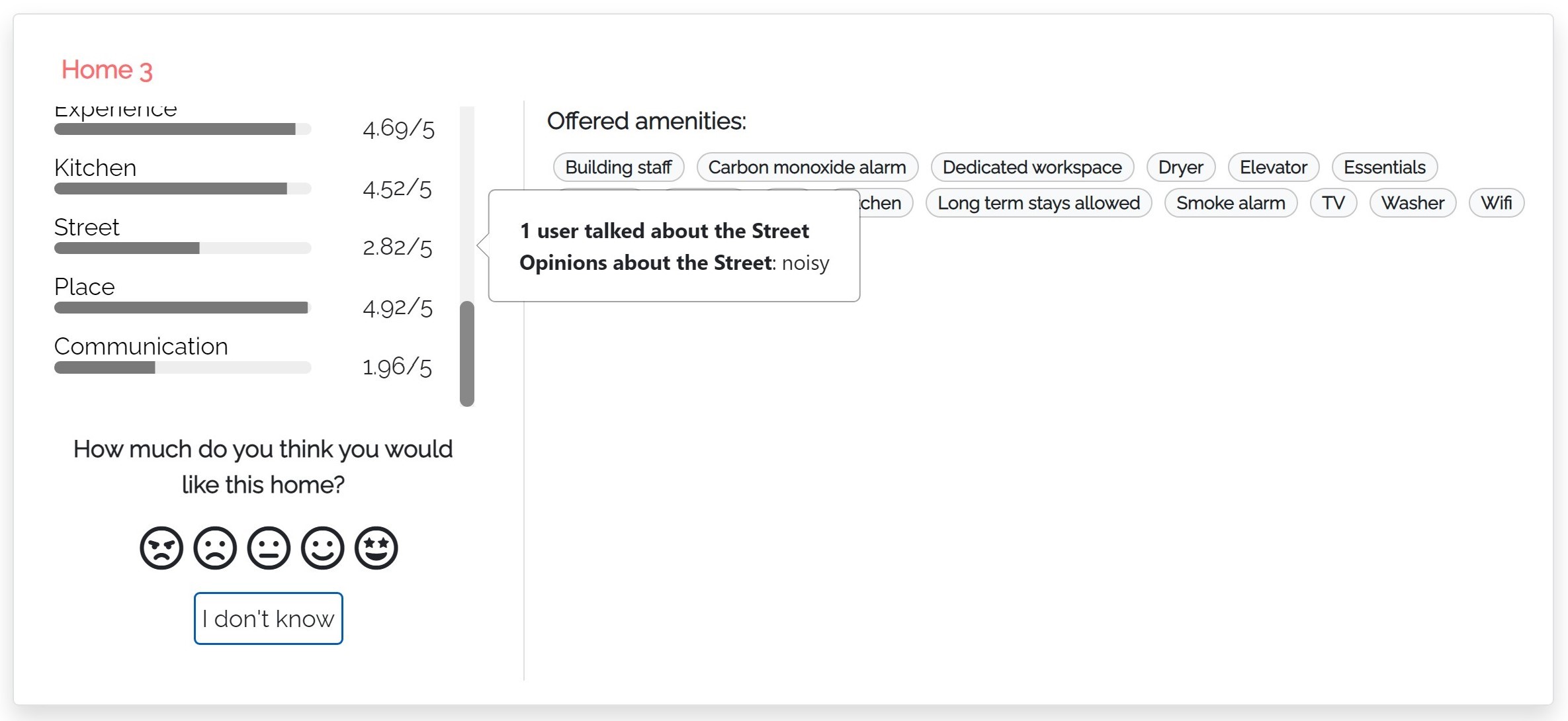}
    \caption{User interface of justification model \modelChen.}
    \label{fig:modelChen-appendix}
\end{sidewaysfigure}





\end{appendices}


\newpage


\begin{thebibliography}{67}
\providecommand{\natexlab}[1]{#1}
\providecommand{\url}[1]{{#1}}
\providecommand{\urlprefix}{URL }
\providecommand{\doi}[1]{\url{https://doi.org/#1}}
\providecommand{\eprint}[2][]{\url{#2}}
 \bibcommenthead

\bibitem[{Abel et~al(2011)Abel, Celik, Houben, and Siehndel}]{Abel-etal:11}
Abel F, Celik I, Houben GJ, et~al (2011) Leveraging the semantics of tweets for
  adaptive faceted search on twitter. In: Aroyo L, Welty C, Alani H, et~al
  (eds) The Semantic Web -- ISWC 2011. Springer Berlin Heidelberg, Berlin,
  Heidelberg, pp 1--17, \doi{10.1007/978-3-642-25073-6\_1},
  \urlprefix\url{https://doi.org/10.1007/978-3-642-25073-6\_1}

\bibitem[{Airbnb(2022)}]{Airbnb}
Airbnb (2022) Airbnb. \url{https://airbnb.com}

\bibitem[{Amal et~al(2019)Amal, Tsai, Brusilovsky, Kuflik, and
  Minkov}]{Amal-etal:19}
Amal S, Tsai CH, Brusilovsky P, et~al (2019) Relational social recommendation:
  application to the academic domain. Expert Systems with Applications 124:182
  -- 195. \doi{10.1016/j.eswa.2019.01.061},
  \urlprefix\url{http://www.sciencedirect.com/science/article/pii/S0957417419300788}

\bibitem[{Bitner et~al(2008)Bitner, Ostrom, and Morgan}]{Bitner-etal:08}
Bitner MJ, Ostrom AL, Morgan FN (2008) Service blueprinting: A practical
  technique for service innovation. California Management Review 50(3):66--94.
  \doi{10.2307/41166446},
  \urlprefix\url{https://doi.org/10.1007/s11257-017-9195-0}

\bibitem[{Cardoso et~al(2019)Cardoso, Sedrakyan, Gutiérrez, Parra,
  Brusilovsky, and Verbert}]{Cardoso-etal:19}
Cardoso B, Sedrakyan G, Gutiérrez F, et~al (2019) {IntersectionExplorer}, a
  multi-perspective approach for exploring recommendations. International
  Journal of Human-Computer Studies 121:73 -- 92.
  \doi{10.1016/j.ijhcs.2018.04.008},
  \urlprefix\url{http://www.sciencedirect.com/science/article/pii/S1071581918301903}

\bibitem[{Chang et~al(2019)Chang, Hahn, Perer, and Kittur}]{Chang-etal:19}
Chang JC, Hahn N, Perer A, et~al (2019) {SearchLens}: composing and capturing
  complex user interests for exploratory search. In: Proceedings of the 24th
  International Conference on Intelligent User Interfaces. ACM, New York, NY,
  USA, IUI '19, pp 498--509, \doi{10.1145/3301275.3302321},
  \urlprefix\url{http://doi.acm.org/10.1145/3301275.3302321}

\bibitem[{Chen and Wang(2017)}]{Chen-Wang:17}
Chen L, Wang F (2017) Explaining recommendations based on feature sentiments in
  product reviews. In: Proceedings of the 22nd International Conference on
  Intelligent User Interfaces. Association for Computing Machinery, New York,
  NY, USA, IUI ’17, p 17–28, \doi{10.1145/3025171.3025173},
  \urlprefix\url{https://doi.org/10.1145/3025171.3025173}

\bibitem[{Chen et~al(2014)Chen, Wang, Qi, and Liang}]{Chen-etal:14}
Chen L, Wang F, Qi L, et~al (2014) Experiment on sentiment embedded comparison
  interface. Knowledge-Based Systems 64:44--58.
  \doi{https://doi.org/10.1016/j.knosys.2014.03.020},
  \urlprefix\url{https://www.sciencedirect.com/science/article/pii/S0950705114001099}

\bibitem[{Chen et~al(2015)Chen, Chen, and Wang}]{Chen-etal:15}
Chen L, Chen G, Wang F (2015) Recommender systems based on user reviews: the
  state of the art. User Modeling and User-Adapted Interaction 25(2):99--154.
  \doi{10.1007/s11257-015-9155-5},
  \urlprefix\url{https://doi.org/10.1007/s11257-015-9155-5}

\bibitem[{Cheng and Jin(2019)}]{Cheng-Jin:19}
Cheng M, Jin X (2019) What do {Airbnb} users care about? {A}n analysis of
  online review comments. International Journal of Hospitality Management 76:58
  -- 70. \doi{https://doi.org/10.1016/j.ijhm.2018.04.004},
  \urlprefix\url{http://www.sciencedirect.com/science/article/pii/S0278431917307491}

\bibitem[{Coelho et~al(2020)Coelho, Hanel, and Wolf}]{Coelho-etal:20}
Coelho G, Hanel PHP, Wolf LJ (2020) The very efficient assessment of need for
  cognition: Developing a six-item version. Assessment 27(8):1870--1885.
  \doi{10.1177/1073191118793208},
  \urlprefix\url{https://doi.org/10.1177/1073191118793208}

\bibitem[{Conati et~al(2021)Conati, Barral, Putnam, and
  Rieger}]{Conati-etal:21}
Conati C, Barral O, Putnam V, et~al (2021) Toward personalized {XAI}: A case
  study in intelligent tutoring systems. Artificial Intelligence 298:103,503.
  \doi{10.1016/j.artint.2021.103503}

\bibitem[{Confalonieri et~al(2021)Confalonieri, Coba, Wagner, and
  Besold}]{Confalonieri-etal:21}
Confalonieri R, Coba L, Wagner B, et~al (2021) A historical perspective of
  explainable artificial intelligence. {WIREs} Data Mining and Knowledge
  Discovery 11(1):e1391. \doi{https://doi.org/10.1002/widm.1391},
  \urlprefix\url{https://wires.onlinelibrary.wiley.com/doi/abs/10.1002/widm.1391},
  {\href{https://arxiv.org/abs/https://wires.onlinelibrary.wiley.com/doi/pdf/10.1002/widm.1391}{{https://arxiv.org/abs/https://wires.onlinelibrary.wiley.com/doi/pdf/10.1002/widm.1391}}}

\bibitem[{Cramer et~al(2008)Cramer, Evers, Ramlal, van Someren, Rutledge,
  Stash, Aroyo, and Wielinga}]{Cramer-etal:08}
Cramer HSM, Evers V, Ramlal S, et~al (2008) The effects of transparency on
  trust in and acceptance of a content-based art recommender. User Modeling and
  User-Adapted Interaction 18(5):455--496. \doi{10.1007/s11257-008-9051-3}

\bibitem[{Di~Noia et~al(2022)Di~Noia, Tintarev, Fatourou, and
  Schedl}]{DiNoia-etal:22}
Di~Noia T, Tintarev N, Fatourou P, et~al (2022) Recommender systems under
  {E}uropean {AI} {R}egulations. Commun ACM 65(4):69–73.
  \doi{10.1145/3512728}, \urlprefix\url{https://doi.org/10.1145/3512728}

\bibitem[{{Di Sciascio} et~al(2016){Di Sciascio}, Sabol, and
  Veas}]{diSciascio-etal:16}
{Di Sciascio} C, Sabol V, Veas EE (2016) Rank as you go: User-driven
  exploration of search results. In: Proceedings of the 21st International
  Conference on Intelligent User Interfaces. Association for Computing
  Machinery, New York, NY, USA, IUI '16, p 118–129,
  \doi{10.1145/2856767.2856797},
  \urlprefix\url{https://doi.org/10.1145/2856767.2856797}

\bibitem[{{Di Sciascio} et~al(2019){Di Sciascio}, Brusilovsky, Trattner, and
  Veas}]{diSciascio-etal:19}
{Di Sciascio} C, Brusilovsky P, Trattner C, et~al (2019) A roadmap to
  user-controllable social exploratory search. ACM Transaction on Interactive
  Intelligent Systems 10(1). \doi{10.1145/3241382},
  \urlprefix\url{https://doi.org/10.1145/3241382}

\bibitem[{Dong and Smyth(2017)}]{Dong-Smyth:17}
Dong R, Smyth B (2017) User-based opinion-based recommendation. In: Proceedings
  26th {IJCAI}, Melbourne, Australia, pp 4821--4825

\bibitem[{{European Commission}(2018)}]{GDPR}
{European Commission} (2018) General data protection regulation ({GDPR)}.
  \url{https://ec.europa.eu/info/law/law-topic/data-protection_en}

\bibitem[{Gibbons(2017)}]{Gibbons:17}
Gibbons S (2017) {Service Blueprints: Definition}.
  \url{https://www.nngroup.com/articles/service-blueprints-definition/}

\bibitem[{Herlocker et~al(2000)Herlocker, Konstan, and
  Riedl}]{Herlocker-etal:00}
Herlocker JL, Konstan JA, Riedl J (2000) Explaining collaborative filtering
  recommendations. In: Proceedings of the 2000 ACM Conference on Computer
  Supported Cooperative Work. Association for Computing Machinery, New York,
  NY, USA, CSCW ’00, p 241–250, \doi{10.1145/358916.358995},
  \urlprefix\url{https://doi.org/10.1145/358916.358995}

\bibitem[{Hern{\'a}ndez-Rubio et~al(2019)Hern{\'a}ndez-Rubio, Cantador, and
  Bellog{\'i}n}]{Rubio-etal:19}
Hern{\'a}ndez-Rubio M, Cantador I, Bellog{\'i}n A (2019) A comparative analysis
  of recommender systems based on item aspect opinions extracted from user
  reviews. User Modeling and User-Adapted Interaction 29(2):381--441.
  \doi{10.1007/s11257-018-9214-9},
  \urlprefix\url{https://doi.org/10.1007/s11257-018-9214-9}

\bibitem[{Hutto and Eric(2014)}]{Hutto-Gilbert:14}
Hutto C, Eric G (2014) {VADER}: A parsimonious rule-based model for sentiment
  analysis of social media text. In: Proceedings of the 8th International
  {AAAI} Conference on Weblogs and Social Media. AAAI, New York, NY, USA, pp
  216--225,
  \urlprefix\url{https://www.aaai.org/ocs/index.php/ICWSM/ICWSM14/paper/viewPaper/8109}

\bibitem[{Jannach et~al(2019)Jannach, Jugovac, and Nunes}]{Jannach-etal:19}
Jannach D, Jugovac M, Nunes I (2019) Explanations and user control in
  recommender systems. In: Proceedings of the 23rd International Workshop on
  Personalization and Recommendation on the Web and Beyond. Association for
  Computing Machinery, New York, NY, USA, ABIS '19, p~31,
  \doi{10.1145/3345002.3349293},
  \urlprefix\url{https://doi.org/10.1145/3345002.3349293}

\bibitem[{Kashdan et~al(2009)Kashdan, Gallagher, Silvia, Winterstein, Breen,
  Terhar, and Steger}]{Kashdan:09}
Kashdan T, Gallagher M, Silvia P, et~al (2009) The curiosity and exploration
  inventory-{II}: Development, factor structure, and psychometrics. Journal of
  research in personality 43:987--998. \doi{10.1016/j.jrp.2009.04.011}

\bibitem[{Kirk(2013)}]{Kirk:13}
Kirk RE (2013) Experimental design: Procedures for the behavioral sciences.
  SAGE Publications, Inc., \doi{10.4135/9781483384733}

\bibitem[{Knijnenburg and Willemsen(2015)}]{Knijnenburg:15}
Knijnenburg BP, Willemsen MC (2015) Evaluating Recommender Systems with User
  Experiments, Springer US, Boston, MA, pp 309--352.
  \doi{10.1007/978-1-4899-7637-6_9},
  \urlprefix\url{https://doi.org/10.1007/978-1-4899-7637-6_9}

\bibitem[{Kouki et~al(2017)Kouki, Schaffer, Pujara, O’Donovan, and
  Getoor}]{Kouki-etal:17}
Kouki P, Schaffer J, Pujara J, et~al (2017) User preferences for hybrid
  explanations. In: Proceedings of the Eleventh ACM Conference on Recommender
  Systems. Association for Computing Machinery, New York, NY, USA, RecSys
  ’17, p 84–88, \doi{10.1145/3109859.3109915},
  \urlprefix\url{https://doi.org/10.1145/3109859.3109915}

\bibitem[{Kouki et~al(2019)Kouki, Schaffer, Pujara, O’Donovan, and
  Getoor}]{Kouki-etal:19}
Kouki P, Schaffer J, Pujara J, et~al (2019) Personalized explanations for
  hybrid recommender systems. In: Proceedings of the 24th International
  Conference on Intelligent User Interfaces. Association for Computing
  Machinery, New York, NY, USA, IUI ’19, p 379–390,
  \doi{10.1145/3301275.3302306},
  \urlprefix\url{https://doi.org/10.1145/3301275.3302306}

\bibitem[{Kouki et~al(2020)Kouki, Schaffer, Pujara, O’Donovan, and
  Getoor}]{Kouki-etal:20}
Kouki P, Schaffer J, Pujara J, et~al (2020) Generating and understanding
  personalized explanations in hybrid recommender systems. ACM Transactions on
  Interactive Intelligent Systems 10(4). \doi{10.1145/3365843},
  \urlprefix\url{https://doi.org/10.1145/3365843}

\bibitem[{Lee(2022)}]{Lee:22}
Lee CKH (2022) How guest-host interactions affect consumer experiences in the
  sharing economy: New evidence from a configurational analysis based on
  consumer reviews. Decision Support Systems 152:113,634.
  \doi{https://doi.org/10.1016/j.dss.2021.113634},
  \urlprefix\url{https://www.sciencedirect.com/science/article/pii/S0167923621001445}

\bibitem[{Lewis and Sauro(2009)}]{Lewis:09}
Lewis JR, Sauro J (2009) The factor structure of the system usability scale.
  In: Kurosu M (ed) Human Centered Design. Springer Berlin Heidelberg, Berlin,
  Heidelberg, pp 94--103

\bibitem[{Loepp et~al(2015)Loepp, Herrmanny, and Ziegler}]{Loepp-etal:15}
Loepp B, Herrmanny K, Ziegler J (2015) Blended recommending: integrating
  interactive information filtering and algorithmic recommender techniques. In:
  Proceedings of the 33rd Annual ACM Conference on Human Factors in Computing
  Systems. ACM, New York, NY, USA, CHI '15, pp 975--984,
  \doi{10.1145/2702123.2702496},
  \urlprefix\url{http://doi.acm.org/10.1145/2702123.2702496}

\bibitem[{Loria(2020)}]{TextBlob}
Loria S (2020) {TextBlob}: Simplified text processing.
  \url{https://textblob.readthedocs.io/en/dev/index.html}

\bibitem[{Lu et~al(2018)Lu, Dong, and Smyth}]{Lu-etal:18b}
Lu Y, Dong R, Smyth B (2018) Why i like it: Multi-task learning for
  recommendation and explanation. In: Proceedings of the 12th ACM Conference on
  Recommender Systems. Association for Computing Machinery, New York, NY, USA,
  RecSys '18, p 4–12, \doi{10.1145/3240323.3240365},
  \urlprefix\url{https://doi.org/10.1145/3240323.3240365}

\bibitem[{Mauro et~al(2020)Mauro, Ardissono, Capecchi, and
  Galioto}]{Mauro-etal:20d}
Mauro N, Ardissono L, Capecchi S, et~al (2020) Service-aware interactive
  presentation of items for decision-making. Applied Sciences, Special Issue
  Implicit and Explicit Human-Computer Interaction 10(16):5599.
  \doi{10.3390/app10165599},
  \urlprefix\url{https://www.mdpi.com/2076-3417/10/16/5599}

\bibitem[{Mauro et~al(2021{\natexlab{a}})Mauro, Hu, and
  Ardissono}]{Mauro-etal:21d}
Mauro N, Hu ZF, Ardissono L (2021{\natexlab{a}}) Service-oriented justification
  of recommender system suggestions. In: Ardito C, Lanzilotti R, Malizia A,
  et~al (eds) Human-Computer-Interaction -- INTERACT 2021, Lecture Notes in
  Computer Science, vol. 12936. Springer International Publishing, Cham,
  Switzerland, pp 321--330, \doi{10.1007/978-3-030-85613-7_23}

\bibitem[{Mauro et~al(2021{\natexlab{b}})Mauro, Hu, Ardissono, and
  Izzi}]{Mauro-etal:21c}
Mauro N, Hu ZFF, Ardissono L, et~al (2021{\natexlab{b}}) A service-oriented
  perspective on the summarization of recommendations: Preliminary experiment.
  In: Adjunct Proceedings of the 29th ACM Conference on User Modeling,
  Adaptation and Personalization. Association for Computing Machinery, New
  York, NY, USA, p 213–219,
  \urlprefix\url{https://doi.org/10.1145/3450614.3464475}

\bibitem[{Mauro et~al(2022)Mauro, Hu, and Ardissono}]{Mauro-etal:22b}
Mauro N, Hu ZF, Ardissono L (2022) Service-aware personalized item
  recommendation. IEEE Access 10:26,715--26,729.
  \doi{10.1109/ACCESS.2022.3157442}

\bibitem[{Millecamp et~al(2019)Millecamp, Htun, Conati, and
  Verbert}]{Millecamp-etal:19}
Millecamp M, Htun NN, Conati C, et~al (2019) To explain or not to explain: the
  effects of personal characteristics when explaining music recommendations.
  In: Proceedings of the 24th International Conference on Intelligent User
  Interfaces. Association for Computing Machinery, New York, NY, USA, IUI
  ’19, p 397–407, \doi{10.1145/3301275.3302313},
  \urlprefix\url{https://doi.org/10.1145/3301275.3302313}

\bibitem[{Millecamp et~al(2020)Millecamp, Htun, Conati, and
  Verbert}]{Millecamp-etal:20}
Millecamp M, Htun NN, Conati C, et~al (2020) What’s in a user? {T}owards
  personalising transparency for music recommender interfaces. In: Proceedings
  of the 28th ACM Conference on User Modeling, Adaptation and Personalization.
  Association for Computing Machinery, New York, NY, USA, UMAP ’20, p
  173–182, \doi{10.1145/3340631.3394844},
  \urlprefix\url{https://doi.org/10.1145/3340631.3394844}

\bibitem[{Millecamp et~al(2022)Millecamp, Conati, and
  Verbert}]{Millecamp-etal:22}
Millecamp M, Conati C, Verbert K (2022) “{K}nowing me, knowing you”:
  personalized explanations for a music recommender system. User Modeling and
  User-Adapted Interaction 32:215--252. \doi{10.1007/s11257-021-09304-9}

\bibitem[{Mohseni et~al(2021)Mohseni, Zarei, and Ragan}]{Mohseni-etal:21}
Mohseni S, Zarei N, Ragan ED (2021) A multidisciplinary survey and framework
  for design and evaluation of explainable {AI} systems. ACM Transactions on
  Interactive Intelligent Systìems 11(3–4). \doi{10.1145/3387166},
  \urlprefix\url{https://doi.org/10.1145/3387166}

\bibitem[{Muhammad et~al(2016)Muhammad, Lawlor, and Smyth}]{Muhammad-etal:16}
Muhammad KI, Lawlor A, Smyth B (2016) A live-user study of opinionated
  explanations for recommender systems. In: Proceedings of the 21st
  International Conference on Intelligent User Interfaces. Association for
  Computing Machinery, New York, NY, USA, IUI '16, p 256–260,
  \doi{10.1145/2856767.2856813},
  \urlprefix\url{https://doi.org/10.1145/2856767.2856813}

\bibitem[{Musto et~al(2019)Musto, Narducci, Lops, {de Gemmis}, and
  Semeraro}]{Musto-etal:19b}
Musto C, Narducci F, Lops P, et~al (2019) Linked open data-based explanations
  for transparent recommender systems. International Journal of Human-Computer
  Studies 121:93 -- 107. \doi{10.1016/j.ijhcs.2018.03.003},
  \urlprefix\url{http://www.sciencedirect.com/science/article/pii/S1071581918300946}

\bibitem[{Musto et~al(2021)Musto, {de Gemmis}, Lops, and
  Semeraro}]{Musto-etal:20}
Musto C, {de Gemmis} M, Lops P, et~al (2021) Generating post hoc review-based
  natural language justifications for recommender systems. User-Modeling and
  User-Adapted Interaction 31:629–673. \doi{10.1007/s11257-020-09270-8}

\bibitem[{Nam et~al(2018)Nam, Kim, and Carnie}]{Nam-etal:18}
Nam KW, Kim BY, Carnie BW (2018) Service open innovation; design elements for
  the food and beverage service business. Journal of Open Innovation:
  Technology, Market, and Complexity 4(4). \doi{10.3390/joitmc4040053},
  \urlprefix\url{https://www.mdpi.com/2199-8531/4/4/53}

\bibitem[{Ni et~al(2019)Ni, Li, and McAuley}]{Ni-etal:19}
Ni J, Li J, McAuley J (2019) Justifying recommendations using distantly-labeled
  reviews and fine-grained aspects. In: Proceedings of the 2019 Conference on
  Empirical Methods in Natural Language Processing and the 9th International
  Joint Conference on Natural Language Processing (EMNLP-IJCNLP). Association
  for Computational Linguistics, Hong Kong, China, pp 188--197,
  \doi{10.18653/v1/D19-1018},
  \urlprefix\url{https://www.aclweb.org/anthology/D19-1018}

\bibitem[{Nunes and Jannach(2017)}]{Nunes-Jannach:17}
Nunes I, Jannach D (2017) A systematic review and taxonomy of explanations in
  decision support and recommender systems. User Modeling and User-Adapted
  Interaction 27(3–5):393–444. \doi{10.1007/s11257-017-9195-0},
  \urlprefix\url{https://doi.org/10.1007/s11257-017-9195-0}

\bibitem[{Parra and Brusilovsky(2015)}]{Parra-Brusilovsky:15}
Parra D, Brusilovsky P (2015) User-controllable personalization: a case study
  with {SetFusion}. International Journal of Human-Computer Studies 78:43 --
  67. \doi{10.1016/j.ijhcs.2015.01.007},
  \urlprefix\url{http://www.sciencedirect.com/science/article/pii/S1071581915000208}

\bibitem[{Pu and Chen(2007)}]{Pu-Chen:07}
Pu P, Chen L (2007) Trust-inspiring explanation interfaces for recommender
  systems. Knowledge-Based Systems 20(6):542 -- 556.
  \doi{10.1016/j.knosys.2007.04.004},
  \urlprefix\url{http://www.sciencedirect.com/science/article/pii/S0950705107000445}

\bibitem[{Pu et~al(2011)Pu, Chen, and Hu}]{Pu-etal:11}
Pu P, Chen L, Hu R (2011) A user-centric evaluation framework for recommender
  systems. In: Proceedings of the Fifth ACM Conference on Recommender Systems.
  Association for Computing Machinery, New York, NY, USA, RecSys ’11, p
  157–164, \doi{10.1145/2043932.2043962},
  \urlprefix\url{https://doi.org/10.1145/2043932.2043962}

\bibitem[{Qiu et~al(2011)Qiu, Liu, Bu, and Chen}]{Qiu:2011}
Qiu G, Liu B, Bu J, et~al (2011) Opinion word expansion and target extraction
  through double propagation. Computational Linguistics 37:9--27.
  \doi{10.1162/coli\_a\_00034}

\bibitem[{Rana et~al(2022)Rana, D’Addio, Manzato, and Bridge}]{Rana-etal:22}
Rana A, D’Addio RM, Manzato MG, et~al (2022) Extended
  recommendation-by-explanation. User-Modeling and User-Adapted Interaction
  32:91–131. \doi{10.1007/s11257-021-09317-4},
  \urlprefix\url{https://doi.org/10.1007/s11257-021-09317-4}

\bibitem[{Ren et~al(2016)Ren, Qiu, Wang, and Lin}]{REN201613}
Ren L, Qiu H, Wang P, et~al (2016) Exploring customer experience with budget
  hotels: Dimensionality and satisfaction. International Journal of Hospitality
  Management 52:13 -- 23. \doi{https://doi.org/10.1016/j.ijhm.2015.09.009},
  \urlprefix\url{http://www.sciencedirect.com/science/article/pii/S0278431915001486}

\bibitem[{Ricci et~al(2022)Ricci, Rokach, and Shapira}]{Ricci-etal:22}
Ricci F, Rokach L, Shapira B (2022) Recommender Systems: Techniques,
  Applications, and Challenges, Springer US, New York, NY, pp 1--35.
  \doi{10.1007/978-1-0716-2197-4_1},
  \urlprefix\url{https://doi.org/10.1007/978-1-0716-2197-4_1}

\bibitem[{Springer and Whittaker(2019)}]{Springer-Whittaker:19}
Springer A, Whittaker S (2019) Progressive disclosure: Empirically motivated
  approaches to designing effective transparency. In: Proceedings of the 24th
  International Conference on Intelligent User Interfaces. Association for
  Computing Machinery, New York, NY, USA, IUI '19, p 107–120,
  \doi{10.1145/3301275.3302322},
  \urlprefix\url{https://doi.org/10.1145/3301275.3302322}

\bibitem[{Stickdorn et~al(2011)Stickdorn, Schneider, and Andrews}]{sdt11}
Stickdorn M, Schneider J, Andrews K (2011) This is service design thinking:
  Basics, tools, cases. Wiley

\bibitem[{Tintarev and Masthoff(2012)}]{Tintarev-Masthoff:12}
Tintarev N, Masthoff J (2012) Evaluating the effectiveness of explanations for
  recommender systems. User Modeling and User-Adapted Interaction
  22(4-5):399--439

\bibitem[{Tintarev and Masthoff(2022)}]{Tintarev-Masthoff:22}
Tintarev N, Masthoff J (2022) Beyond Explaining Single Item Recommendations,
  Springer US, New York, NY, pp 711--756. \doi{10.1007/978-1-0716-2197-4_19},
  \urlprefix\url{https://doi.org/10.1007/978-1-0716-2197-4_19}

\bibitem[{TripAdvisor(2017)}]{TripAdvisor}
TripAdvisor (2017) Tripadvisor. \url{https://www.tripadvisor.it/}

\bibitem[{Tsai and Brusilovsky(2019)}]{Tsai-Brusilovsky:19}
Tsai CH, Brusilovsky P (2019) Exploring social recommendations with visual
  diversity-promoting interfaces. ACM Transactions on Interactive Intelligent
  Systems 10(1):5:1--5:34. \doi{10.1145/3231465},
  \urlprefix\url{http://doi.acm.org/10.1145/3231465}

\bibitem[{Tsai and Brusilovsky(2021)}]{Tsai-Brusilovsky:21}
Tsai CH, Brusilovsky P (2021) The effects of controllability and explainability
  in a social recommender system. User Modeling and User-Adapted Interaction
  31(3):591--627. \doi{10.1007/s11257-020-09281-5}

\bibitem[{Tvaro\v{z}ek et~al(2008)Tvaro\v{z}ek, Barla, Frivolt, Tom\v{s}a, and
  Bielikov\'{a}}]{Tvarozek-etal:08}
Tvaro\v{z}ek M, Barla M, Frivolt G, et~al (2008) Improving semantic search via
  integrated personalized faceted and visual graph navigation. In: Proceedings
  of the 34th Conference on Current Trends in Theory and Practice of Computer
  Science. Springer-Verlag, Berlin, Heidelberg, SOFSEM'08, pp 778--789,
  \doi{10.1007/978-3-540-77566-9\_67},
  \urlprefix\url{http://dl.acm.org/citation.cfm?id=1785934.1786006}

\bibitem[{Verbert et~al(2016)Verbert, Parra, and Brusilovsky}]{Vebert-etal:16}
Verbert K, Parra D, Brusilovsky P (2016) Agents vs. users: visual
  recommendation of research talks with multiple dimension of relevance. ACM
  Transactions on Interactive Intelligent Systems 6(2). \doi{10.1145/2946794},
  \urlprefix\url{https://doi.org/10.1145/2946794}

\bibitem[{Wang et~al(2018)Wang, Zhang, Wang, Zhao, Li, Xie, and
  Guo}]{Wang-etal:18}
Wang H, Zhang F, Wang J, et~al (2018) Ripplenet: propagating user preferences
  on the knowledge graph for recommender systems. In: Proceedings of the 27th
  ACM International Conference on Information and Knowledge Management.
  Association for Computing Machinery, New York, NY, USA, CIKM ’18, p
  417–426, \doi{10.1145/3269206.3271739},
  \urlprefix\url{https://doi.org/10.1145/3269206.3271739}

\bibitem[{Yi et~al(2020)Yi, Yuan, and Yoo}]{Yi-etal:20}
Yi J, Yuan G, Yoo C (2020) The effect of the perceived risk on the adoption of
  the sharing economy in the tourism industry: The case of {Airbnb}.
  Information Processing \& Management 57(1):102,108.
  \doi{https://doi.org/10.1016/j.ipm.2019.102108},
  \urlprefix\url{https://www.sciencedirect.com/science/article/pii/S0306457319301347}

\end{thebibliography}


\end{document}